\documentclass[onecolumn,amsmath,amssymb,11pt]{revtex4} 

\usepackage{graphicx}
\usepackage{dcolumn}
\usepackage{bm}
\input epsf

\newcommand{\nn}{\nonumber}
\newcommand{\be}{\begin{equation}}
\newcommand{\ee}{\end{equation}}
\newcommand{\bea}{\begin{eqnarray}}
\newcommand{\eea}{\end{eqnarray}}

\usepackage[pdftex]{hyperref}

\def\b{\alpha}
\def\d{\delta}
\def\e{\epsilon}
\def\p{\phi}

\def\ps{\psi}
\def\th{\theta}
\def\r{\rho}
\def\s{\sigma}
\def\t{\tau}
\def\z{\zeta}
\def\half{\frac{1}{2}}
\def\htip{h_{{\rm tip}}}
\def\btip{b_{{\rm tip}}}
\def\tht{\Xi}

\def\pc{\p_{c}}
\def\T{T_{\sigma s}}
\def\eend{\epsilon_e}
\def\es{\epsilon_*}
\def\cse{c_{se}}
\def\css{c_{s*}}
\def\ns{n_s}

\def\Fthree{f_{NL}^{(3)}}
\def\Floc{f_{NL}^{loc}}
\def\Feq{f_{NL}^{eq}}
\def\dfe{\d \p_e}
\def\rb{\beta}
\def\nNG{n_{{\rm NG} }}

\def\sNL{s_{NL}}
\def\tNL{t_{NL}}
\def\CT{{\cal T}}
\def\Tn{T_{loc,\;eq}}
\def\Tz{T_\zeta^{{\rm \, loc,\,eq}}}
\def\zs{z_s}

\def\la{\langle}
\def\ra{\rangle}
\def\picube{(2\pi)^3}
\def\bkone{\mathbf k_1}
\def\bktwo{\mathbf k_2}
\def\bkthree{\mathbf k_3}
\def\bkfour{\mathbf k_4}

\def\bk{{\bf k}}
\def\bkp{{\bf k'}}

\def\fNL{{f_{NL}}}

\def\fNLl{f_{NL}^{1\;\rm{loop}}}
\def\Prat{\frac{P_{\zeta}^{1loop}}{P_{\zeta}^{tree}}}
\def\Pz{\mathcal{P}_{\zeta}}

\newcommand{\MP}{M_{\rm pl}}
\newcommand{\sdelta}[1]{\!\delta^{\,3}(\mathbf{#1})}

\begin{document}

\title{Combined local and equilateral non-Gaussianities from multifield DBI inflation}

\author{S\'{e}bastien Renaux-Petel\footnote{renaux@apc.univ-paris7.fr}}
\affiliation{ APC (UMR 7164, CNRS, Universit\'e Paris 7), 10 rue Alice Domon et L\'eonie Duquet, 75205 Paris Cedex 13, France
}

\date{\today}

\begin{abstract}

We study multifield aspects of Dirac-Born-Infeld (DBI) inflation. More specifically, we consider an inflationary phase driven by the radial motion of a D-brane in a conical throat and determine how the D-brane fluctuations in the angular directions can be converted into curvature perturbations when the tachyonic instability arises at the end of inflation. The simultaneous presence of multiple fields and non-standard kinetic terms gives both local and equilateral shapes for non-Gaussianities in the bispectrum. We also study the trispectrum, pointing out that it acquires a particular momentum dependent component whose amplitude is given by $f_{NL}^{loc} \, f_{NL}^{eq}$. We show that this relation is valid in every multifield DBI model, in particular for any brane trajectory, and thus constitutes an interesting observational signature of such scenarios. 

\end{abstract}

\maketitle

\tableofcontents{}

\section{Introduction}

Current measurements of the cosmic microwave background (CMB) anisotropies, such as those obtained by the WMAP satellite, already provide us a with a wealth of valuable information about the very early universe. Furthermore, with the successful launch of the Planck satellite and the increasing precision of large scale structure surveys, one can hope to get yet more precise information in the near future. In this context non-Gaussianity \cite{Komatsu:2009kd} is particularly exciting since it has the ability to discriminate between models which are otherwise degenerate at the linear level: for instance, a detection of so-called local non-Gaussianity would rule out all single field scenarios of inflation in a model independent way \cite{Creminelli:2004yq}. Amongst such single field models, those with standard kinetic terms, in which the inflaton slowly rolls down its potential, come with an unobservably low level of non-Gaussianity and hence are still consistent with present-day observations \cite{Komatsu:2008hk}. Theoretically though, their embedding in high-energy physics theories is hampered by the eta-problem, namely that Planck-suppressed corrections often lead to potentials that are too steep to support slow-roll inflation. In the context of string theory, this problem was demonstrated to be particularly acute in slow-roll inflationary models based on the dynamics of D-branes moving in higher dimensional spaces \cite{Baumann:2007ah,Baumann:2007np}. Nonetheless, this set-up precisely motivates an interesting way to bypass the eta-problem that has attracted a lot of attention: since the inflaton field in brane inflation is governed by a Dirac-Born-Infeld (DBI) action characterized by non-canonical kinetic terms, there exists an upper bound on the inflaton velocity that allows one to achieve an inflationary phase with otherwise too steep potentials \cite{Silverstein:2003hf}. Besides alleviating the eta-problem, the non standard kinetic terms also enhance the self-interactions of the inflaton, resulting in significant non-Gaussianities of equilateral type \cite{Alishahiha:2004eh}. Indeed, the simplest single-field DBI models are already under strain from observations \cite{Bean:2007hc,Peiris:2007gz}. 

\smallskip

However, as is central to the discussion of this paper, DBI inflation is naturally a multifield scenario, since the position of the brane in each compact extra dimension gives rise to a scalar field from the effective four-dimensional point of view \cite{Kehagias:1999vr,Steer:2002ng,Brax:2002jj,Easson:2007fz}. In multiple field models, the scalar perturbations can be decomposed into (instantaneous) adiabatic and entropy modes by projecting, respectively, parallel and perpendicular to the background trajectory in field space \cite{Gordon:2000hv}. If the entropy fields are light enough to be quantum mechanically excited during inflation, they develop super-Hubble fluctuations that can be transferred to the adiabatic mode on large scales. This effect, as well as the nonlinearities imprinted on the fields at the epoch of horizon crossing, was taken into account in \cite{Langlois:2008wt} where it was shown that the shape of \textit{equilateral} non-Gaussianities is the same as in the single-field case while their amplitude is reduced by the entropy to curvature transfer, which therefore eases the confrontation with the data. This paper, as well as subsequent ones on multifield DBI inflation \cite{Langlois:2008qf,Arroja:2008yy,Langlois:2009ej,Gao:2009gd,Gao:2009fx,Mizuno:2009cv,Gao:2009at}, focused mainly on equilateral non-Gaussianities. More generally, multiple-field inflationary models are known to produce possibly large non-Gaussianities of another shape, namely \textit{local} non-Gaussianities, that arise due to the nonlinear classical evolution of perturbations on superhorizon scales. The formalism developed in these papers also remained very general and no particular model was presented for the entropic transfer on large scales. We address both these questions here, using a mechanism outlined by Lyth and Riotto \cite{Lyth:2006nx} to convert entropic into adiabatic perturbations at the end of brane inflation.
 
\smallskip

In this scenario, inflation is still driven by a single inflaton scalar field, namely the D3-brane solely moves along the standard radial direction of the throat in the context of warped conical compactifications. When the mobile D3-brane and an anti D3-brane sitting at the tip of the throat come within a string length, an open string mode stretched between them becomes tachyonic, triggering their annihilation and the end of inflation \cite{Sen:2002in}. As the brane-antibrane distance is \textit{six-dimensional}, it acquires some dependence upon the fluctuations of the light fields parametrizing the angular position of the brane. Hence the value of the inflaton at which the instability signals is modulated and the duration of inflation varies from one super-Hubble region to another. In this way initially entropic perturbations are converted into the curvature perturbation. The relevance of this mechanism in the slow-roll 'Delicate Universe' scenario \cite{Baumann:2007ah} has been recently investigated \cite{Chen:2008ada}. Here however we would like to combine it with the DBI inflationary regime. This was looked at by Leblond and Shandera \cite{Leblond:2006cc} regarding the power spectrum. In the present paper, we extend their analysis by taking into account the enhancement of the angular fluctuations by the low speed of sound \cite{Langlois:2008wt}, as well as by investigating the non-Gaussian properties of the curvature perturbation. It should be noted that while no explicit model of DBI inflation in a string theory framework has been constructed yet that both satisfy observational constraints and are theoretically self-consistent \footnote{We are considering the so-called ultra-violet model of DBI inflation, in which the mobile D3-brane falls towards the tip of the throat, but consistent infrared models \cite{Chen:2004gc,Chen:2005ad} can be constructed, as explained in \cite{Bean:2007eh}.} -- mainly because of the existence of a geometrical limit for the size of the throat in Planckian units \cite{Baumann:2006cd,Lidsey:2007gq}\footnote{This bound was first discussed in \cite{Chen:2006hs} in the context of eternal D-brane inflation.} -- it was so far assumed that the fluctuations of the primary inflaton create the curvature perturbation. One can therefore hope to embed consistently the DBI inflationary scenario in string theory with other mechanisms to generate the density perturbations, such as the one considered in this paper.

In our model, the curvature perturbation is nonlinearly related to the entropy perturbations, therefore the conversion process gives rise to local non-Gaussianities besides the standard equilateral ones generated at horizon crossing. To the best of our knowledge it is the first time that definite predictions are made for an inflationary scenario where both significant local and equilateral non-Gaussianities can arise, characterized in the bispectrum respectively by the parameters $\Floc$ and $\Feq$. While the logical possibility that a linear combination of different shapes of the bispectrum can arise is an acknowledged fact, it should be stressed that what we consider is not merely a juxtaposition of an inflaton with non standard kinetic terms generating $\Feq$ and another light scalar generating $\Floc$. In multifield DBI inflation, light scalar fields other than the inflaton \textit{and} with derivative interactions naturally contribute to both types of non-Gaussianities. We show that this nontrivial combination leaves a distinct imprint on the primordial \textit{trispectrum}, which acquires a particular momentum dependent component whose amplitude is given by the product $\Floc\, \Feq$. This relation is structural and is valid independently of the details of the inflationary scenario, {\it i.e.} for any brane trajectory and any process by which entropic perturbations feed the adiabatic ones. Hence it constitutes an interesting observational signature of multifield DBI inflation.

 \bigskip
 
The layout of this paper is the following. In section \ref{sec:end}, we describe our set-up and recall results concerning the amplification of quantum fluctuations in multifield DBI inflation. We explain the mechanism by which entropic perturbations are converted into the curvature perturbation at the end of brane inflation. Using the $\delta N$ formalism, we also derive the relevant formulae for quantifying this effect. In section \ref{sec:obs}, we calculate the power spectrum of the primordial curvature perturbation. We show, in particular, that the entropic transfer is more efficient in the DBI than in the slow-roll regime. We also calculate the primordial bispectrum which acquires a linear combination of both the local and equilateral shapes of non-Gaussianities. We then combine our results for the spectrum and bispectrum to derive constraints on the model. In section \ref{trispectrum} we turn to the study of the primordial trispectrum. We calculate the local trispectrum parameters $\tau_{NL}$ and $g_{NL}$ and discuss the purely quantum contribution coming from the field trispectra. Then we point out the presence of a particular component of the trispectrum whose amplitude is given by the product $\Floc\, \Feq$. We finish by plotting the corresponding shape of the trispectrum in different limits, and this turns out to have characteristic features. We summarize our main results in the last section.

\section{Generating the curvature perturbation at the end of brane inflation}
\label{sec:end}

\subsection{The set-up and the amplification of quantum fluctuations}

Our setting is that of a flux compactification of type IIB string theory to four dimensions \cite{Giddings:2001yu}, resulting in a warped geometry in which the six-dimensional Calabi-Yau manifold contains one or more throats. The ten-dimensional metric inside a throat has the generic form
\be
ds^2 = h^{2}(y^K)\,g_{\mu \nu}dx^\mu dx^\nu + h^{-2}(y^K)\, G_{IJ}(y^K)\, dy^I dy^J \,,
\label{metric}
\ee 
where $g_{\mu \nu}$ is the metric of the four-dimensional, non-compact, spacetime and we have factored out the so-called warp-factor $h(y^I)$ from the metric $G_{IJ}$ in the six compact extra dimensions. In the following, we assume that the geometry presents a special radial direction, in agreement with known solutions of the supergravity equations \cite{Klebanov:2000hb}, so that the warp factor is a function of a radial coordinate $\r$ only, decreasing along the throat down to its tip at $\r=0$ \footnote{We use the symbol $\r$ to evade any ambiguities with the commonly used variable $r$ in the Klebanov-Strassler throat \cite{Klebanov:2000hb} that has a non-zero minimum value, although the two coincide far from the tip in the KS throat.}. In this framework, we consider the following internal metric
\bea
G_{IJ}(y^K)\, dy^I dy^J &=&d \r^2 + b^2(\r)\, g_{m n}^{(5)} d \ps^m d \ps^n \, , 
\label{internal_metric}
\eea
where we refer to $b(\r)$ as the radius of the throat and $\ps^m$ ($m$\,=\,5,6,7,8,9) denote its angular coordinates. Since we aim to present a general mechanism, we do not specify a precise form for $h(\r)$ and $b(\r)$ and only require that they approach constant values $\htip$ and $\btip$ near the tip, which is the situation encountered for instance in the Klebanov-Strassler throat \cite{Klebanov:2000hb}.

We take then a probe D3-brane, of tension
\be
T_3=\frac{m_s^4}{(2\pi)^3 g_s}\,,
\label{T3}
\ee
where $m_s$ is the string mass and $g_s$ the string coupling, filling the four-dimensional spacetime, and point like in the six extra dimensions. The D3-brane has coordinates $y^I_{(b)}$ and falls down to the tip of the throat where a static $\overline{D3}$ is sitting. Following \cite{Langlois:2008wt}, in terms of the rescaled scalar fields
\be
\p^I = \sqrt{T_3}\,y^I_{(b)} \to \p=\sqrt{T_3}\, \r_{(b)} \,, \th^m=\sqrt{T_3}\, \ps^m_{(b)}\,,
\label{rescaling}
\ee
and rescaled warp factor
\be
f=\left(T_3 h^4 \right)^{-1}\,,
\label{f-h}
\ee
the D3-brane low-energy dynamics is captured by the Lagrangian
\be
P= -f(\p)^{-1}\left(\sqrt{ \det(\delta^{\mu}_{\nu}+f \, G_{IJ}\partial^{\mu} \p^I \partial_{\nu} \p^J )}-1\right) -V(\p^I)\,,
\label{DD}
\ee
where $V(\p^I)$ is the field interaction potential. Note that in general, there are also contributions from the presence of various $p$-forms in the bulk as well as the gauge field confined on the brane. In \cite{Langlois:2009ej}, these fields were shown to have no observable effects on scalar cosmological perturbations at least to second-order so we have omitted them here for simplicity.

The explicit calculation of the potential $V$ in (\ref{DD}) is extremely difficult and requires a detailed knowledge of the compactification scheme (see e.g \cite{Baumann:2009ni} and references therein). In general though, we know that bulk as well as moduli stabilizing effects break the isometries of the throat, stabilizing some of the angular coordinates of the branes. However, there typically remain approximate residual isometries of the potential, as shown explicitly in \cite{Chen:2008ada} for the most-advanced brane inflation model \cite{Baumann:2007np}. For simplicity, we consider only one such isometry direction $\ps$, entering the five-dimensional metric $g_{mn}^{(5)}$ (\ref{internal_metric}) through
\be
g_{m n}^{(5)} d \ps^m d \ps^n =  d \ps^2 + \ldots
\label{gmn}
\ee
and we assume that the four other brane angular degrees of freedom are frozen in their minima of their effective potential at the position of the antibrane along these directions. Therefore we take the potential  $V=V(\p)$ to depend only on the radial position of the brane, which itself moves along the radial direction only -- $\dot \p \neq 0\,, \dot \th^m =0$. As opposed to single-field inflation, the perturbations along the isometric, {\it i.e.} flat direction $\ps$, can be quantum mechanically excited during inflation. In that case, the angular $D3-\overline{D3}$ separation $\th \equiv \sqrt{T_3} \Delta \psi$ varies from one Hubble patch to the others. Although this does not modify the dynamics \textit{during} inflation, this will turn out to be crucial \textit{at the end} of inflation, as we will see below.

We now recall the relevant results of \cite{Langlois:2008qf} regarding the amplification of quantum fluctuations in multifield brane inflation, in particular the amplitude of the inflaton and of the angular perturbations at horizon crossing. For that purpose, it is convenient, after going to conformal time $\tau = \int {{\rm d}t}/{a(t)}$, where $a(t)$ is the cosmological scale factor, to work in terms of the canonically normalized fields given by
\be
v_{\s}=\frac{a}{c_s^{3/2}} \, Q_{\p} \,,\qquad \,v_{s}=\frac{a}{\sqrt{c_s}}\, b(\p)\,Q_{\th}\,,
\label{v}
\ee
where $Q^I$ denotes the perturbations of the field $\p^I$ in the flat gauge and
\bea
c_s &\equiv& \sqrt{1-f(\p) \dot \p^2}
\label{cs}
\eea
is the propagation speed of scalar perturbations (see Eq.~(\ref{eq_v}) below), or speed of sound. Clearly, from Eq.~(\ref{cs}), there is an upper bound on the inflaton velocity $ |  \dot \p | \leq \frac{1}{\sqrt{f(\p)}}$. When $c_s^2 \approx 1$, one can expand the square-root in the Lagrangian (\ref{DD}) to quadratic order in the fields. Then the action becomes canonical and one recovers the slow-roll regime. However, when the brane almost saturates its speed limit -- $c_s^2 \ll 1$ -- the non-standard structure of the action (\ref{DD}) must be fully taken into account: this is the relativistic, or DBI regime we are particularly interested in. 

In Fourier space the equations of motion for $v_\s$ and $v_s$ at linear order take the simple form \cite{Langlois:2008mn,Langlois:2008qf}
\begin{eqnarray}
v_{\s}''+\left(c_s^2 k^2-\frac{z''}{z}\right) v_{\s} =0\,, \qquad
v_{s}''+\left(c_s^2 k^2- \frac{\zs''}{\zs}+a^2\mu_s^2\right) v_{s}=0\,,
\label{eq_v}
\end{eqnarray}
where we have introduced the two background-dependent functions $z(\t)=a \sqrt{2 \epsilon} /c_s\,, \zs(\t)=a/\sqrt{c_s}$, with $\epsilon \equiv -\frac{\dot H}{H^2}$ the inflationary deceleration parameter. The effective entropic mass squared $\mu_s^2$ is given by
\begin{eqnarray}
\mu_s^2&\equiv&c_s\frac{b'(\p)}{b(\p)}  V'(\p) -\frac{(1-c_s)^2}{2f^2}\frac{b'(\p)}{b(\p)}f'(\p)-\dot \p^2 \frac{b''(\p)}{b(\p)}\,,
\label{mus2}
\end{eqnarray}
where $b(\p)$ is the radius of the throat evaluated at the brane position. In the following we assume that the time evolution of  $\epsilon$ and $c_s$ is very slow with respect to that of the scale factor \footnote{See \cite{Khoury:2008wj} for models where the speed of sound is rapidly varying.}, as quantified by the slow-varying parameters
\bea
\eta&\equiv&\frac{\dot \e}{H \e} \ll 1\,, \label{eta} \\
s&\equiv&\frac{\dot c_s}{H c_s} \ll 1 \label{s} \,,
\eea
so that $z''/z\simeq \zs''/\zs \simeq 2/\tau^2$. The amplification of the vacuum fluctuations at horizon crossing is possible only for very light degrees of freedom. Although this is automatically verified for the adiabatic perturbation $v_{\s}$ because of our assumption $z''/z\simeq 2/\tau^2$, this is not necessary true for $v_s$; if $\mu_s^2$ is larger than $H^2$, this amplification is suppressed and there is no production of entropy modes. Note that the effective entropic mass squared (\ref{mus2}) is non zero despite the angular direction being exactly isometric. In particular, because the potential and radius of the throat typically increase and $f$ typically decreases with $\p$, the first two terms in (\ref{mus2}) are positive. Below we assume that $|\mu_s^2|/H^2\ll 1$ so that the entropy modes are effectively amplified.

Following the standard procedure (see e.g. \cite{Mukhanov:1990me} or \cite{Langlois:2004de}), one then selects the positive frequency solutions of Eq.~(\ref{eq_v}), which  correspond to the Minkowski-like vacuum on very small scales:
  \be
v_{\s\, k} \simeq v_{s\, k} \simeq  \frac{1}{\sqrt{2k c_s}}e^{-ik c_s \tau }\left(1-\frac{i}{k c_s\tau}\right)\, .
\ee
As a consequence, the power spectra for $v_\s$ and $v_s$ after sound horizon crossing  have the same amplitude. However, in terms of the initial field perturbations, one finds, using (\ref{v}), 
\be
{\cal P}_{Q_\p*}\simeq  \left( \frac{H_*}{2 \pi } \right)^2, \quad {\cal P}_{b_*Q_{\th*}}\simeq  \left(\frac{H_*}{2 \pi c_{s*}}  \right)^2
\label{enhancement}
\ee
(the subscript $*$ indicates that the corresponding quantity is evaluated at sound horizon crossing $k c_s=aH$). Therefore, for small $\css$, the entropic modes are {\it amplified} with respect to the adiabatic modes. As we will discuss in subsection \ref{subsec:deltaN}, the standard formulae of the $\delta N$ formalism are expressed in terms of fields whose perturbations have the canonical amplitude $H_*/2 \pi$ at horizon crossing. We therefore define the 'canonical' entropy field as
\be
\tht \equiv b_* c_{s *} \th \,.
\label{can}
\ee

\subsection{The conversion process}

\begin{figure}[t]
\begin{center}
\includegraphics[width=0.6\textwidth]{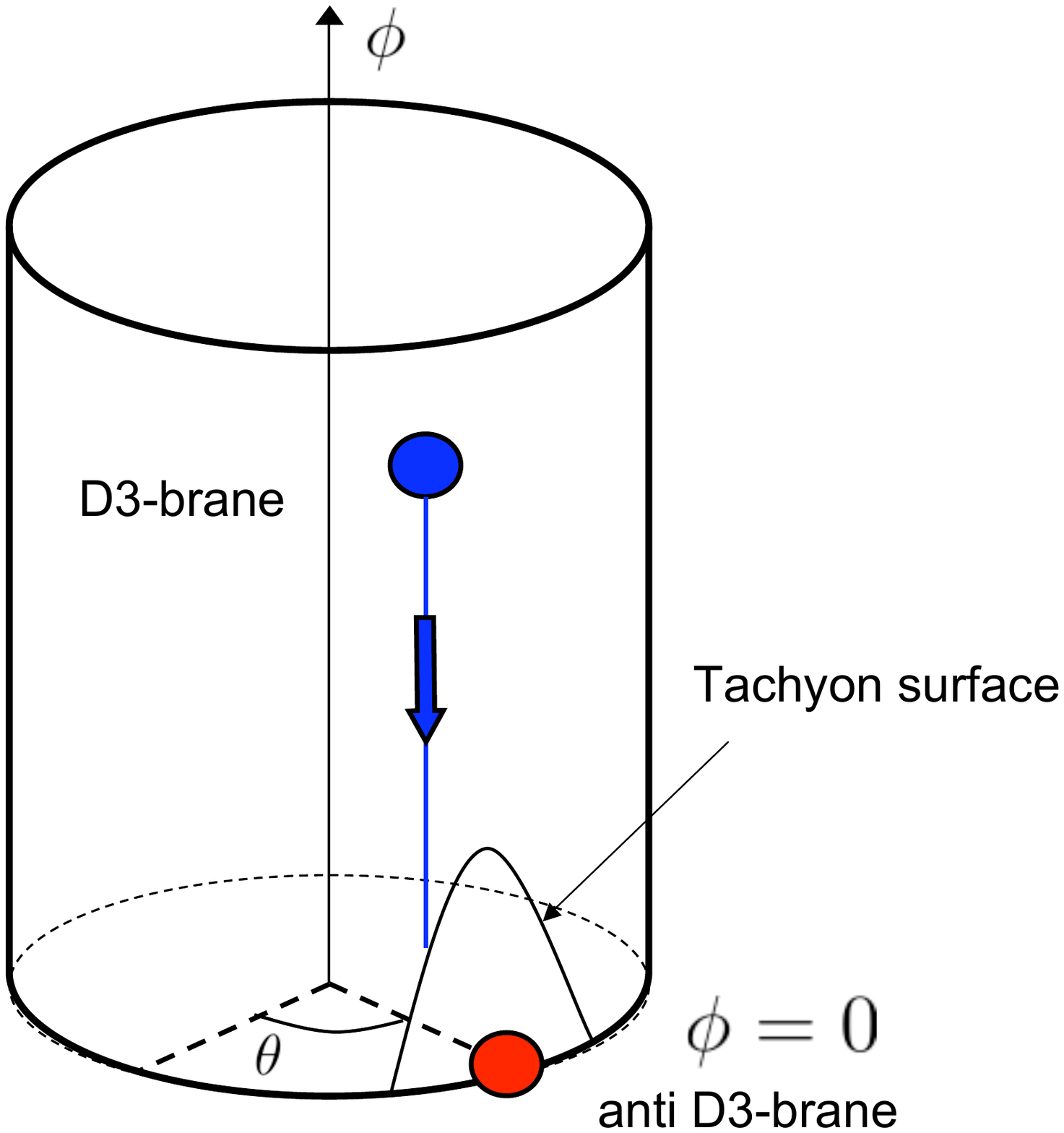}
\caption{\label{cylindre} {\small A simplified picture of the geometry at the tip of the throat, with one angular direction $\th$ only. The radial inflationary trajectory is represented by the blue line.
}}
\end{center}
\end{figure}

\begin{figure}[t]
\begin{center}
\includegraphics[width=0.6\textwidth]{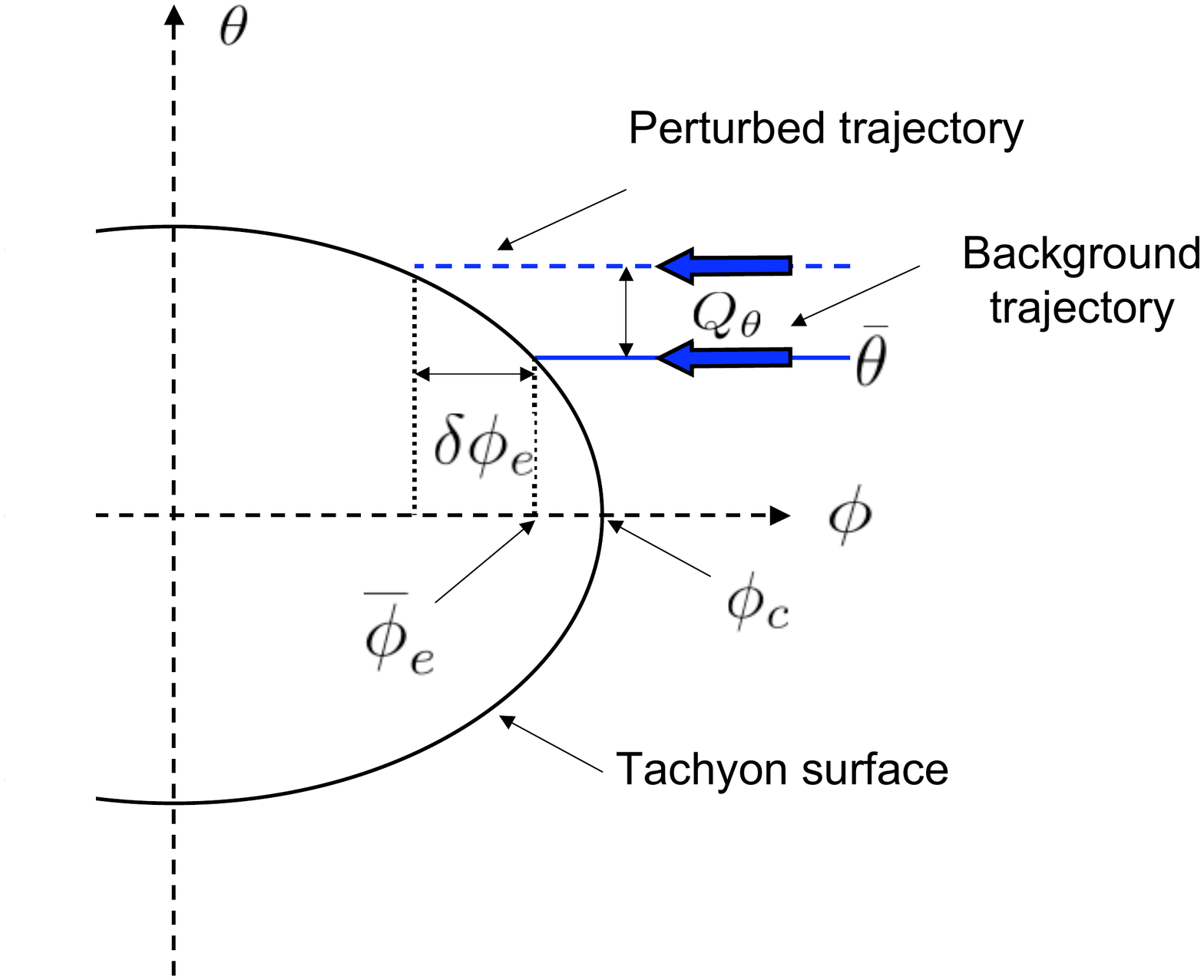}
\caption{\label{trajectoire} {\small The tachyon surface, at which inflation ends, in the $\phi-\th$ plane. The end value of the inflaton is shifted from ${\overline \p_e}$ to ${\overline \p_e} +\delta \p_e$ due to the angular fluctuation $Q_{\th}$, hence the duration of inflation varies from one super-Hubble region to another.}}
\end{center}
\end{figure}

We now address the question of how initially entropic perturbations can be converted into the adiabatic modes. We assume that inflation does not end not by the breakdown of the slow-roll conditions but rather persists all along down the throat. Then, when the D3-brane comes within a string length of the anti D3-brane, a tachyonic instability arises which ends inflation. Using Eqs.~(\ref{metric}), (\ref{internal_metric}) and (\ref{gmn}), this happens when
\be
\frac{1}{\htip^2}\left(( \Delta \r)^2+\btip^2 (\Delta \ps)^2 \right)=l_s^2 \,,
\ee
where $ \Delta \r$ and $\Delta \ps$ are the radial and angular $D3-\overline{D3}$ separation and $l_s=m_s^{-1}$ is the string length. In terms of the rescaled fields, the tachyon surface, represented in Figs.~(\ref{cylindre}) and (\ref{trajectoire}), is given by \footnote{Notice that for the KS throat, its metric is often described with another radial variable, $\t$ \cite{Klebanov:2000hb}, such that $\r \sim \sqrt{g_s M} \htip l_s \t$ near the tip, where $M \gg 1$ is a flux integer. With this variable, the tachyon appears -- forgetting about the angular direction -- for $\t \sim 1/\sqrt{g_s M}$, where the warped string length $\htip l_s$ does not appear. This field however is not canonically normalized and, had we used this variable, the warped string length $\htip l_s$ would reappear in the amplitude of the corresponding perturbation.}
\be
\p^2+\btip^2 \th^2= T_3\, l_s^2 \, \htip^2 \,.
\label{end-inflation}
\ee
The main point is that the end value of the inflaton acquires a spatial dependence through the fluctuations of the light angular field $\th$ (see Fig.~(\ref{trajectoire})). Consequently, the duration of inflation varies from one super-Hubble region to another and this can be interpreted as a curvature perturbation, as we will quantify in the next subsection. To simplify the notation, we define
\be
\pc \equiv \sqrt{T_3} l_s \htip = \frac{m_s \htip}{(2\pi)^{3/2} g_s^{1/2}}
\label{pc}
\ee
and the angle $-\pi/2<\b< \pi/2$ such that the background value of the inflaton field when it reaches the tachyon surface ${\overline \p_e} $ is given by
\be
{\overline \p_e} = \cos(\b)\, \pc\,.
\label{alpha}
\ee
(here and in the following, the subscript $e$ indicates the end of infation). For example, the $D3-\overline{D3}$ angular separation vanishes when $\b=0$.

Let us now make some remarks regarding the validity of our scenario: clearly the $D3$-brane will reach the tachyon surface during its radial fall-down only if the background angular brane separation $\overline{\Delta \ps}$ verifies
\be
\overline{\Delta \ps} \leq  \frac{l_s \htip}{\btip}\,.
\label{reach-tachyon}
\ee
For the supergravity approximation to be valid, the radius at the tip $\btip$ must be large in local string units $\htip l_s$ (it is $\sim \sqrt{g_s M}$ in the KS throat with $M \gg 1$ a flux integer). Therefore, condition Eq.~(\ref{reach-tachyon}) is a non-trivial requirement, which we assumed to be fulfilled. Note however that if this is not satisfied, inflation could end through the angular motion of the brane at the tip \cite{DeWolfe:2004qx,DeWolfe:2007hd,Pajer:2008uy}. The angular fluctuations would then still produce a spatially-dependent time-delay to the end of inflation, and one could expect similar effects to the ones discussed below to arise. Finally, we have also assumed for simplicity that the trajectory is completely radial until the end of inflation: more generally the trajectory can be nontrivial in the angular directions, for instance if the Coulombic attraction between the branes becomes important towards the end of inflation. In that case, the entropic perturbations would feed the curvature perturbation through the bending of the inflationary trajectory \cite{Gordon:2000hv,Bernardeau:2002jy}. This mechanism to convert entropic into adiabatic perturbations is different from the one used in this paper but both can be present, as in multi-brid inflation \cite{Sasaki:2008uc,Naruko:2008sq,Byrnes:2008zy,Huang:2009xa}\footnote{The effect of the bending of the trajectory on non-Gaussianities can be analyzed with the results of \cite{RenauxPetel:2008gi,Lehners:2009ja}, based on the earlier work \cite{Langlois:2006vv}, in particular when it is difficult to apply the $\delta N$ formalism.}

\subsection{$\delta N$ formulae}
\label{subsec:deltaN}

In order to quantify the curvature perturbation generated by the entropic fluctuations, it is convenient to use the $\d N$ formalism \cite{Starobinsky:1986fx,Sasaki:1995aw,Sasaki:1998ug,Lyth:2004gb,Lyth:2005fi}, in which a key role is played by the local integrated expansion, or local number of e-folds, between some initial and final hypersurfaces
\be
N({\bf x})=\int_{i}^{f} H(t,{\bf x}) d t\,.
\ee
In this formalism, the curvature perturbation on uniform energy density hypersurfaces, which we denote $\zeta$, is identified as the perturbation in the local  number of e-folds of expansion from an initially flat hypersurface to a final uniform energy density hypersurface
\be
\z=\d N \equiv N({\bf x})-\bar N\,,
\ee
where $\bar N$ is the number of e-folds in the homogeneous background spacetime. In the long-wavelength limit, according to the separate universe picture \cite{Wands:2000dp}, the integrated expansion can be calculated from solutions to the \textit{unperturbed} Friedmann equation, with initial conditions specified by the \textit{perturbed} scalar fields $\zeta= N(\varphi^A_*)- \bar N$, where $\varphi^A_*=\overline{ \varphi^A_*} + Q^A_*$ is the sum of the homogeneous values plus fluctuations of the scalar fields on the initial spatially-flat slice, which we take to be soon after horizon crossing during inflation. Taylor-expanding this relation in terms of the field fluctuations leads to the formal expression
\be
\z=N_A Q^A +\half N_{AB}Q^A Q^B +\frac{1}{6}N_{ABC} Q^A Q^B Q^C+\ldots
\label{expansion}
\ee
Once $N(\varphi^A_*)$ is known, one can work out the coefficients $N_A,N_{AB}, N_{ABC}$ in (\ref{expansion}) and determine the curvature perturbation.

In our model, since the background dynamics is solely determined by the radial scalar field $\p$, the local number of e-folds, evaluated right after the end of inflation, simply reads
\be
N(\p_*( {\bf x}),\tht_*({\bf x}))=\int_{\p_*}^{\p_e(\tht_*)}\left( \frac{H}{\dot \p}\right) \, d \p \,,
\label{N}
\ee
where the canonical entropy field $\Xi$ was defined in (\ref{can}) and, using Eqs.~(\ref{end-inflation}) and (\ref{pc}), the end value of the inflaton reads
\be
\p_e(\tht_*)=\sqrt{\pc^2-\left(\frac{\rb}{c_{s*}} \tht_*\right)^2}\,,
\label{pe}
\ee
with
\be
\rb \equiv  \frac{\btip}{b_*} 
\label{beta}
\ee
being the ratio between the radius at the tip of the throat and at sound horizon crossing. Here we assumed that no significant expansion is generated after the field reaches the tachyon surface, that is, the sudden end approximation \cite{Lyth:2005qk}. If not negligible, the extra curvature perturbation generated can be taken into account similarly to \cite{Sasaki:2008uc}, though even at this stage, $\zeta$ may not have settled down to its final value. For example, the further evolution of the tachyon can give other contributions to $\zeta$ \cite{Barnaby:2006km,Enqvist:2005nc}. This interesting aspect is outside the scope of this paper but it should be borne in mind that it is present in general.

From Eq.~(\ref{N}), one obtains
\be
\z=  \z_{*}+\z_e
\label{zeta}
\ee
where
\be
\z_{*}= -\int_{\bar{\p}_*}^{\bar{\p}_*+Q_{\p *}}\left( \frac{H}{\dot \p}\right) \, d \p = - \left. \left(\frac{H}{\dot \p} \right) \right|_{*}  Q_{\p *}-  \left. \half \frac{{\rm d}}{{\rm d}\p}\left(\frac{H}{\dot \p} \right) \right|_{*}  Q_{\p *}^2 
-  \left. \frac{1}{3!} \frac{{\rm d^2}}{{\rm d}\p^2}\left(\frac{H}{\dot \p} \right) \right|_{*}  Q_{\p *}^3+ \ldots
\label{zeta*}
\ee
and
\be
\z_e= \int_{{\bar\p}_e( \bar{\tht})}^{\p_e( \bar{\tht}+Q_{\tht *} )}\left( \frac{H}{\dot \p}\right)\, d \p= \left. \left(\frac{H}{\dot \p} \right) \right|_{e}  \dfe +\left. \half \frac{{\rm d}}{{\rm d}\p}\left(\frac{H}{\dot \p} \right) \right|_{e}  \dfe^2 
+  \left. \frac{1}{3!} \frac{{\rm d^2}}{{\rm d}\p^2}\left(\frac{H}{\dot \p} \right) \right|_{e}  \dfe^3+ \ldots
\label{zetae}
\ee
are the contributions to the curvature perturbation from respectively, the epoch of horizon crossing and the end of inflation. Loosely speaking, $\z_e$ is associated to the time delay generated by the fluctuation $\d \p_e$ in Fig.~(\ref{trajectoire}), where we use the notation
\bea
\dfe = \p_e' Q_{\tht_*}  +\half \p_e^{(2)} (Q_{\tht_*})^2 +\frac{1}{3!}\p_e^{(3)}(Q_{\tht_*})^3 + \ldots
\label{main-nl}
\eea
and the derivatives of $\p_e(\tht_*)$ (\ref{pe}), given in Appendix 1, are evaluated on the background. Note that in order to trust the perturbative expansion, the angular fluctuations must be small compared to the background separation, namely $\rb H_*/\css \ll  \p_c$.

Adding (\ref{zeta*}) and (\ref{zetae}) gives an expression of $\zeta$ of the form (\ref{expansion}), where $A,B= \s, s$ with $Q_{\s} \equiv -Q_{\p_*}$ and $Q_{s}\equiv Q_{\tht_*}$, which are normalized to share the canonical amplitude $H_*/2 \pi$. To leading order in the slow-varying parameters $\eta$ and $s$ (\ref{eta})-(\ref{s}) and their time derivatives, only the first terms remain in  the expansions (\ref{zeta*}) and (\ref{zetae}) so that the non-zero coefficients in (\ref{expansion}) are then
\bea
&N_{\s}&=- \left.  \frac{1}{\sqrt{2 \e c_{s}}}\frac{1}{\MP} \right|_{*}   \label{Nsi} \qquad N_s=- \left. \frac{\p_e'}{\sqrt{2 \e c_{s}}}\frac{1}{\MP}   \right|_{e}   \label{Ns} \\
&N_{ss}&=-\left.\frac{\p_e^{(2)}}{\sqrt{2 \e c_s}\MP}    \right|_{e}  \label{Nss} \qquad N_{sss}= - \left. \frac{\p_e^{(3)}}{\sqrt{2 \e c_s}\MP} \right|_{e} \,  \label{Nsss}
\eea
where we have used $H/\dot \p=-(2 \e c_s)^{-1/2}/\MP$ \cite{Langlois:2008qf}. For completeness, we include in Appendix 1 the full expansion of $\zeta$, not restricting to leading order in the slow-varying parameters. Notice also that they cannot be neglected in the computation of the scalar spectral index $n_s$ and running non-Gaussian parameter $\nNG$ as they give then the leading order result (see below).

\section{Power spectrum and primordial non-Gaussianities from the bispectrum}
\label{sec:obs}

According to Eq.~(\ref{expansion}), the statistical properties of the curvature perturbation $\z$ are inherited from those of the field fluctuations $Q^A$ at horizon crossing. For clarity, we first recall the results, determined in \cite{Langlois:2008qf}, for the field two-point and three-point functions in two-field DBI inflation, before calculating the power spectrum and bispectrum of $\zeta$ in the following subsections. The investigation of the trispectrum will be the subject of section \ref{trispectrum}. We use the notations of \cite{Byrnes:2006vq}.

\subsection{Statistical properties of the field perturbations at horizon crossing}
\label{subsec:fields-correlators}

In Fourier space the power spectrum of the scalar field perturbations is defined by
\be
\la Q_\bk^A Q_\bkp^B \ra = C^{AB}(k)\picube \,  \sdelta{\bk+\bkp} \,.
\label{def-CAB}
\ee
To leading order in the field perturbations 
\begin{equation}
C^{AB}(k) = \frac{H_*^2}{2k^3} \delta^{AB}\,,
\label{CAB}
\end{equation}
where $\delta^{AB}$ is the Kronecker delta-function. Notice that the cross-correlation between adiabatic and entropy modes is zero for the straight line background trajectory considered here as the coupling between them exactly vanishes in that case (see Eq.~(\ref{eq_v}) as well as \cite{Langlois:2008qf} for details).

The bispectrum of the field perturbations is defined by
\begin{equation}
\la Q^A_{{\mathbf k_1}}\,Q^B_{{\mathbf k_2}}\, Q^C_{{\mathbf k_3}}  \ra
\equiv B^{ABC}(k_1,k_2,k_3) \picube \sdelta{\bkone+\bktwo+\bkthree}\,.
\label{defbispectrum}
\end{equation}
In slow-roll inflation, where the self interactions of the fields are suppressed by the flatness of the potential, the bispectrum of the fields is small, both for single \cite{Maldacena:2002vr} and multifield inflation \cite{Seery:2005gb}. On the contrary, self-interactions are enhanced in models with non-standard kinetic terms \cite{Chen:2006nt}, and in DBI inflation in the low sound speed limit in particular, with the result \cite{Langlois:2008qf}
\bea 
B^{ABC}(k_1,k_2,k_3) = \frac{H_{*}^4}{4 \sqrt{2 \e_* c_{s*} }c_{s*}^2 \MP}  b^{ABC}(k_1,k_2,k_3)\,. 
\label{three-point}
\eea
The fully adiabatic momentum dependent factor $ b^{ABC}(k_1,k_2,k_3)$ is given by
\bea
b^{\s \s \s}( k_1,k_2,k_3)&=&\frac{1}{\prod_i k_i^3 K^3}  \left[ 6 k_1^2 k_2^2 k_3^2
- k_3^2 (\bk_1 \cdot \bk_2)(2 k_1 k_2 -k_3 K +2 K^2)  + {\rm perm.}\right]
 \label{b-adiabatic}
\eea
where $K=k_1+k_2+k_3$ and the `perm.' indicate two other terms with the same structure as the last term but permutations of indices 1, 2 and 3. Note also that it depends only on the norm of the three wave-vectors as for instance, $\bk_1 \cdot \bk_2 =\half (k_3^2-k_1^2-k_2^2)$ due to momentum conservation. This is the standard result from single field DBI inflation \cite{Chen:2005fe}. In the relativistic limit, there exists only one other non-zero three-point correlation function at leading order, namely
\bea
b^{\s s s}( k_1,k_2,k_3) &=&\frac{1}{\prod_i k_i^3 K^3} 
\left[ 2 k_1^2 k_2^2 k_3^2+ k_1^2 (\bk_2 \cdot \bk_3)(2 k_2 k_3 -k_1 K +2 K^2) \right.
\cr
&-&\left.  k_3^2 (\bk_1 \cdot \bk_2)(2 k_1 k_2 -k_3 K +2 K^2) 
-k_2^2 (\bk_1 \cdot \bk_3)(2 k_1 k_3 -k_2 K +2 K^2) \right].
\label{b-entropic}
\eea

\subsection{Power spectrum, scalar spectral index and tensor to scalar ratio}
\label{General results}

The power spectrum of the curvature perturbation is defined as
\begin{equation}
\la \zeta_\bk \zeta_\bkp \ra = P_\zeta(k)\picube \sdelta{\bk+\bkp} \,.
\label{defP}
\end{equation}
The corresponding variance per logarithmic interval in $k$-space is given, to leading order in the field perturbations, by
\begin{equation}
{\cal P_{\z}}(k) \equiv \frac{k^3}{2\pi^2} P_{\z}(k)=\left(\frac{H_*}{2\pi}\right)^2 \left( N_{\s}^2+N_s^2 \right) \,,
\label{curvature-general}
\end{equation}
where we have used Eqs.~(\ref{expansion}) and (\ref{CAB}). As in \cite{Wands:2002bn,Langlois:2008qf}, it is convenient to introduce the transfer function $\T$, such that $N_s= \T N_{\s}$. The curvature power-spectrum then takes the form
\be
{\cal P_{\z}} = \frac{1}{8 \pi^2 \e_* c_{s*}} \frac{H_*^2}{\MP^2} \left( 1+\T^2 \right) \,,
\label{curvature}
\ee
where $\T^2$ quantifies the contribution of the entropy modes to the final curvature perturbation. This vanishes in single-field DBI inflation while in our case, from Eq.~(\ref{Ns}),
\be
\T^2=\frac{\es}{\eend \cse\css} \tan^2(\b)\, \rb^2 \,.
\label{transfer}
\ee
Let us comment on the ranges of the various parameters entering Eq.~(\ref{transfer}). First, since the radius of the throat decreases from the UV to the IR end, $\rb \equiv \btip/b_* $ is bounded by one and this bound is saturated when the last 60 efolds of inflation take place at the tip of the throat, where $b(\p)$ becomes a constant \cite{Kecskemeti:2006cg}. Second, the entropic transfer depends on the angular $D3-\overline{D3}$ separation through the angle $\b$ (\ref{alpha}). When $\tan(\b)=0$, the angular fluctuations give no time-delay to the end of inflation at linear order, as is clear from Fig.~(\ref{trajectoire}), and hence the transfer function vanishes. In that case, the first effect appears through higher order loop corrections (see Appendix 2). In the following, we have in mind that $\tan(\b)=O(1)$ although we keep formulae general. Finally, the result (\ref{transfer}) indicates that in slow-roll inflation, where the speed of sound is one, the entropic contribution to the curvature power spectrum can be significant, compared to the inflaton one, only if the deceleration parameter $\e$ is smaller at the end of inflation than at horizon crossing. In DBI inflation, however, the transfer function is amplified by the inverse of the product of the sound speed at horizon crossing and at the end of inflation, hence it is more efficient. Note that the enhancement of entropic perturbations by the inverse of the sound speed (\ref{enhancement}) was crucial in deriving this result.

It is straightforward to calculate the scalar spectral index and tensor to scalar ratio from the power spectrum. We find
\be
\ns -1\equiv \frac{{\rm d\,ln}{\cal P_{\z}}}{{\rm d\, ln\,}k}=-2\epsilon_*- \eta_* -s_* +\frac{\T^2}{1+\T^2}\left(\eta_*-s_*-2 \frac{\dot b_*}{H_* b_*} \right)\,,
\label{spectral-index}
\ee
and \cite{Langlois:2008qf}
\be
r=16 \es \css \frac{1}{1+\T^2} \approx 16 \frac{1}{\tan^2(\b)} \frac{1}{\rb^2 }\eend \cse \css^2\,, \qquad \T^2 \gg 1
\ee
where the last equality holds in the limit of a large entropic transfer. Hence it is clear that, when the curvature perturbation is of entropic origin, the links between the observables and the microscopic parameters of the model are completely different from the single field case. 

\subsection{Primordial bispectrum}

\subsubsection{General definitions}

The bispectrum of the curvature perturbation is defined as
\begin{equation}
\langle\zeta_{{\mathbf k_1}}\,\zeta_{{\mathbf k_2}}\, \zeta_{{\mathbf k_3}}\rangle \equiv
B_\zeta( k_1,k_2,k_3) \picube \sdelta{{\mathbf k_1}+{\mathbf k_2}+{\mathbf k_3}} \,,
\label{3-point}
\end{equation}
where, from Eqs.~(\ref{expansion}), (\ref{def-CAB}) and (\ref{defbispectrum}) and to leading order \cite{Allen:2005ye} 
\bea 
 && \hspace*{-1.5em} B_\zeta(k_1,k_2,k_3) = N_A N_B N_C B^{ABC}(k_1,k_2,k_3) \nn \\
&+& N_A N_{BC} N_D \left[ C^{AC}(k_1) C^{BD}(k_2) + C^{AC}(k_2) C^{BD}(k_3)+ C^{AC}(k_3)
C^{BD}(k_1) \right] \,.
 \label{zetabispectrum} 
\eea
Observational quantities are usually expressed in terms of the dimensionless non-linearity parameter $\fNL$ -- generally momentum-dependent -- defined by
\bea
B_\zeta( k_1,k_2,k_3)  &=&
 \frac65 \fNL \left[ P_\zeta(k_1) P_\zeta(k_2)  + {\rm perm.} \right] \,.
 \label{fNL-def}
 \eea
 Hence there are two contributions to $\fNL$: the first, coming from the first term in Eq.~(\ref{zetabispectrum}), is related to the three-point functions of the fields at horizon crossing, and we will denote it as $\Fthree$:
\bea
 \Fthree = \frac{5}{6} \frac{N_A N_B N_C B^{ABC}(k_1,k_2,k_3)}{\left( P_\zeta(k_1)P_\zeta(k_2)+  {\rm perm.}\right) } \,.
  \label{fNLeq}
\eea
The second, coming from the second group of terms in Eq.~(\ref{zetabispectrum}), comes from the leading order nonlinear relation between the curvature perturbation and the field perturbations, and we will denote it as $\Floc$ \cite{Lyth:2005fi}:
\bea
\Floc= \frac{5}{6} \frac{N_A N_B N^{AB}}{\left(N_C N^C\right)^2} \,,
  \label{fNLloc}
\eea
the total $\fNL$ being the sum of the two $\fNL=\Fthree+\Floc$.

\subsubsection{Equilateral and local bispectra}

Let us first discuss $\Fthree$: if inflation is of slow-roll type when the observables modes cross the horizon -- $\css^2 \approx 1$ -- then $\Fthree$ is negligibly small. We therefore concentrate on the relativistic regime $\css^2 \ll 1$, in which case (\ref{three-point}), (\ref{b-adiabatic}), (\ref{b-entropic}) and (\ref{curvature-general}) give \cite{Langlois:2008wt}
\be
 \Fthree = -\frac{5}{6 \css^2 (1+\T^2)} \left( b^{\s \s \s}( k_1,k_2,k_3) \frac{\prod_i k_i^3}{\sum_i k_i^3} \right)  \,.
  \label{fNL-our-model}
\ee
Here we used the symmetry property $ b^{\s s s}( k_1,k_2,k_3)+ b^{s \s s}( k_1,k_2,k_3)+ b^{s s \s}( k_1,k_2,k_3)= b^{\s \s \s}( k_1,k_2,k_3)$, which implies that the shape dependence of $\Fthree$ is the same as in single-field DBI, as can be understood in a geometrical way \cite{Mizuno:2009cv}. In the equilateral limit $k_1=k_2=k_3$,
\be
\Feq =-\frac{35}{108}\frac{1}{\css^2}\frac{1}{1+\T^2 }\,,
\label{f_NL3}
\ee
where $\T^2$ is given in (\ref{transfer}) in our model. Hence the entropic to curvature transfer in multifield DBI inflation diminishes the amount of equilateral non-Gaussianities with respect to the single field case. This comes from the fact that the transfer not only enhances the bispectrum of $\z$ but it also enhances its power spectrum by the same amount. Since $\Feq$ is roughly the ratio of the three-point function with respect to the square of the power spectrum, $\Feq$ is effectively reduced.

We now turn to the local shape of the bispectrum: from the definition (\ref{fNLloc}), we obtain \footnote{Note that strictly speaking there exists also a purely adiabatic contribution to $\Floc$, proportional to $N_{\s \s}$ (\ref{Nsisi}). This equals $\frac{5 (\eta_* + s_*)}{12 (1+\T^2)^2}$, so it is suppressed by both slow-varying parameters and by the entropic transfer and is hence unobservably small. Similar, purely adiabatic, contributions are present in the non-linearity parameters of the trispectrum and we will neglect them as well.}
\be
\Floc=\frac{5}{6}\frac{  N_s^2 N_{s s}}{\left(N_{\s}^2+N_s^2\right)^2}\,.
\label{Floc-general}
\ee
When there is a large entropic transfer -- $N_s^2 \gg N_{\s}^2$ -- (the case where the entropic transfer is small is discussed in Appendix 2) Eq.~(\ref{Floc-general}) reduces to the single field \textit{entropic} result
\be
f_{NL }^{loc}=\frac{5}{6}\frac{N_{ss}}{N_s^2} = -\frac{5}{3} \sqrt{\frac{\eend \cse}{2}} \frac{\p_e^{(2)} \MP}{\p_e'^2}\,, \qquad \T^2 \gg 1\,,
\label{FNL4-again}
\ee
where we have used Eqs.~(\ref{Ns})-(\ref{Nss}) in the last equality. Up to the factor $\sqrt{\cse}$ which diminishes the effect, this expression is identical to the slow-roll result \cite{Lyth:2006nx} in the same limit. We have however a concrete model at hand, for which one obtains
\be
f_{NL }^{loc}=\frac{5}{3}\frac{1}{\sin^2(\b) \cos(\b)} \sqrt{\frac{\eend \cse}{2}} \frac{\MP}{\pc}\,, \qquad \T^2 \gg 1\,.
\label{FNL4-iso}
\ee
This shows that $\Floc$ is always positive, in agreement with the discussion in \cite{Huang:2009vk}, and can be significant, given that the inflaton field $\p$ is highly sub-Planckian in brane inflation \cite{Baumann:2006cd} and therefore that $\frac{\MP}{\pc} \gg 1$. As we will see in subsection \ref{sec:implication}, it can even saturate the existing observational bound and put constraints on the model. 

\bigskip

Let us stress that the bispectrum in our scenario displays a combination of two different shapes \footnote{Note that this could be realized in the different context of multiple branes inflation \cite{Cai:2008if,Cai:2009hw}.}. First, the classical nonlinear relation between the curvature perturbation and the light entropic scalar field gives rise to local non-Gaussianities, that peak for squeezed triangles ($k_3 \ll k_2 \approx k_1$), and this is quantified by the parameter $\Floc$. This type of non-Gaussianities constantly arises when the curvature perturbation is generated at the end of inflation through light fields other than the inflaton \cite{Bernardeau:2002jf,Dvali:2003em,Kofman:2003nx,Zaldarriaga:2003my,Bernardeau:2004zz,Lyth:2005qk,Salem:2005nd,Alabidi:2006wa,Bernardeau:2007xi,Dutta:2008if}. Second, derivative interactions produce quantum correlations at the epoch of horizon crossing between modes of comparable wavelengths. The associated non-Gaussian signal peaks for equilateral triangles in momentum space ($k_1 \sim k_2 \sim k_3$) and this is quantified by the parameter $\Feq$. Note that as the local and equilateral signals have fairly orthogonal distributions in momentum space \cite{Babich:2004gb}, observational bounds on each of them can be used when they are both present, each one being almost blind to the other \footnote{We thank Eiichiro Komatsu for pointing this to us. Note also the recent paper \cite{Senatore:2009gt} which demonstrates that the analysis of the shapes of non-Gaussianities requires careful handling.}.

\subsubsection{Running non-Gaussianities}

Besides its amplitude and shape, the scale dependence of the primordial bispectrum is another probe of the early universe physics, and recently it has been shown that combining CMB and large scale structure observations give interesting constraints on the running of non-Gaussianities \cite{LoVerde:2007ri,Sefusatti:2009xu}. In our scenario, while $\Floc$ (\ref{FNL4-iso}) is scale-independent, it is clear that $\Feq$ (\ref{f_NL3}) is scale-dependent. This can be quantified by the running non-Gaussian parameter, defined as
\be
\nNG  \equiv \frac{\partial \, {\rm ln}   | \Feq(k) |  }{\partial \, {\rm ln} k}= -2s_* - \frac{\T^2}{1+\T^2}\left(\eta_*-s_*-2 \frac{\dot b_*}{H_* b_*} \right)\,,
\ee
where we have used Eqs.~(\ref{transfer}) and (\ref{f_NL3}) in the second equality.

As the speed of sound generally decreases with time in models in which the brane goes from the UV to the IR end of the throat, $s\equiv \dot c_s / H c_s < 0 $ so that $\nNG$ is positive if the entropy modes do not feed the curvature perturbation. On the contrary, using the relation $\eta =2\epsilon -s -\frac{\dot f}{Hf}$ valid in the DBI regime (see \cite{Langlois:2009ej} for details), one obtains, for a large transfer,
\be
\nNG = -2 \e_* +2 \frac{\dot b_*}{H_* b_*} +  \frac{\dot f_*}{H_* f_*} \,, \qquad \T^2 \gg 1\,.
\ee
Hence, if observable modes cross the horizon at the tip of the throat where $b$ and $f$ become constant, the running non-Gaussian index rather becomes negative. 

\subsection{Implication of the results}
\label{sec:implication}

In the recent paper \cite{Bird:2009pq}, it was noticed that in some models, brane inflation ends by tachyonic instability in the relativistic DBI regime, even if inflation is of slow-roll type when the observable modes cross the horizon. In this subsection, we concentrate on this limit, namely $\cse^2 \ll 1$, and additionally assume that the curvature perturbation is mostly of entropic origin $\T^2 \gg 1$. Note that the deceleration parameter $\e$ has been used until now only as a small parameter, quantifying how much the inflationary expansion is close to de-Sitter. However, in the DBI regime, it can be related to the brane tension and the warp factor. Indeed, since
\be
\e \equiv -\frac{\dot H}{H^2}=\frac{\dot \p^2}{2c_s H^2 \MP^2}\,,
\ee
when $c_s^2 =1-f \dot \p^2 \ll 1 $, this gives, with Eq.~(\ref{f-h}),
\be
\e c_s = \frac{T_3 h^4}{2 H^2 \MP^2} \,.
\label{e-dbi}
\ee
Therefore, from (\ref{transfer}) and (\ref{e-dbi}), our hypotheses imply the condition
\bea
\frac{2H_e^2 \MP^2}{T_3 \htip^4}\frac{\es}{\css}\tan^2(\b) \,\rb^2 \gg 1 
\eea
on the parameters of the model. With these assumptions, we now combine the results of the previous subsections and discuss the constraints imposed by non-Gaussianities from the bispectrum.

From $\Floc$ given in (\ref{FNL4-iso}), together with the definitions of $\pc$ in Eq.~(\ref{pc}) and $T_3$ in Eq.~(\ref{T3}), we obtain
\be
f_{NL}^{loc}=\frac{5}{6}\frac{1}{\sin^2(\b) \cos(\b)} \frac{\htip m_s}{H_e}\,.
\label{FNL4-simplified}
\ee
To avoid stringy corrections, one requires that at least $\frac{\htip m_s}{H_e} \gtrsim 1$ \cite{Frey:2005jk}, which therefore acts in the direction of making observable local non-Gaussianities \footnote{A more severe bound can even be derived if one requires that the four-dimensional energy-density $\rho_e$ be less than the local string energy density $(\htip m_s)^4$. With the Friedmann equation $3 \MP^2 H_e^2 = \rho_e$, this gives indeed $\htip m_s/H_e \gtrsim \left( \frac{\MP}{H_e}\right)^{1/2} \gtrsim 140$ where the last inequality follows from the non detection of tensor modes. As this is only an order of magnitude bound, we do not consider that it rules out the model.}. This can be made more stringent by noting that the power spectrum (\ref{curvature}) can be reexpressed, in the limit of a large entropic transfer, as
\be
{\cal P_{\z}} =  2 \pi g_s\,  \tan^2(\b)\,\left(\frac{\rb}{\css}\right)^2 \, \frac{(H_* H_e)^2}{(m_s \htip)^4}\,.
\label{curvature-entropic-simplified}
\ee
Then, Eqs.~(\ref{FNL4-simplified}) and (\ref{curvature-entropic-simplified}) together lead to
\be
f_{NL}^{loc}=\frac{5}{6}  2^{3/2}\, \frac{(2 \pi g_s)^{1/4}}{\sin^{3/2}(2 \b)} \, \frac{1}{{\cal P_{\z}}^{1/4}} \,\left( \frac{\rb}{\css} \right)^{1/2} \, \left(\frac{H_*}{H_e}\right)^{1/2}     \,.
\label{FNL4-full-simplified}
\ee
For instance, taking $g_s=0.1$ and $\b=\pi/4$ -- which minimizes the result -- together with the observed normalization of the power spectrum ${\cal P_{\z}}=2.41 \,. 10^{-9}$ \cite{Komatsu:2008hk} gives
\be
f_{NL}^{loc} \simeq 300 \left(\frac{\rb}{\css} \right)^{1/2}  \left(\frac{H_*}{H_e}\right)^{1/2}     \,.
\label{FNL4-estimate}
\ee
Let us recall that the only hypotheses that we used to derive (\ref{FNL4-estimate}) are that the curvature perturbation is of entropic origin and that the end of infation takes place in the relativistic regime. It is in particular valid for any value of $\css$ between zero and one. Furthermore, since the Hubble scale decreases during inflation, namely $H_*/H_e > 1$, then we find the lower bound
\be
f_{NL}^{loc}  \geq 300\,  \rb^{1/2}\,.
\ee
Therefore, the WMAP5 observational constraint $-9 < f_{NL}^{loc} < 111\,\,(95 \% \, {\rm CL})$ \cite{Komatsu:2008hk} implies the upper bound $\rb\equiv \btip/b_* \lesssim 0.1$ in order to avoid too large non-Gaussianities of local type.

We now turn to equilateral non-Gaussianities, which are diluted by the entropic transfer (\ref{f_NL3})
\be
\Feq \simeq-0.3\frac{1}{\css^2 \T^2}=-0.3 \frac{1}{\tan^2(\b)} \frac{\eend \cse}{\es \css} \frac{1}{\rb^2} \,.
\label{f_NLeq}
\ee
With Eq.~(\ref{e-dbi}), this gives
\be
\Feq \simeq-0.3 \frac{1}{\tan^2(\b)}   \left(\frac{H_*}{H_e}\right)^{2}  \frac{1}{\rb^2}   \left(\frac{\htip}{h_*}\right)^{4} \,.
\label{f_NLeq-dbi}
\ee
One needs a precise model to actually determine the amplitude of $\Feq$. Let us simply comment that if one tunes $\rb \equiv \frac{\btip}{b_*}$ to a small value to low down $f_{NL}^{loc}$ (\ref{FNL4-estimate}) in the observational range, this enhances $f_{NL}^{eq}$. However, in that case, observable modes cross the horizon far from the tip of the throat, and one expects a huge hierarchy $  \left(\frac{\htip}{h_*}\right)^{4}  \ll 1$, which tends to put $f_{NL}^{eq}$ within the current observational bounds $-151< \Feq < 253\,\,   (95 \% \, {\rm CL})$ \cite{Komatsu:2008hk}.

\section{Primordial non-Gaussianities from the trispectrum}
\label{trispectrum}

As the next generation of experiments will be able to probe refined details of the statistics of density fluctuations \cite{Kogo:2006kh,Cooray:2008eb,Jeong:2009vd}, the study of the primordial trispectrum is becoming increasingly important. In DBI inflation, another motivation comes from the fact that $\Fthree$ acquires the same momentum dependence in single- and multiple-field models. Therefore, we cannot observationally differentiate between them with the bispectrum alone. As the degeneracy between models tends to be broken as we go to higher-order correlation functions, the investigation of the trispectrum in multifield DBI inflation is thus very natural.

\subsection{General definitions}

The primordial trispectrum is defined as the connected part of the four-point correlation function of the curvature perturbation in Fourier space
\be
\la \zeta_{{\mathbf k_1}}\,\zeta_{{\mathbf k_2}}\, \zeta_{{\mathbf k_3}} \zeta_{{\mathbf
k_4}} \ra_c \equiv T_\zeta({\mathbf k_1},{\mathbf k_2},{\mathbf k_3}, {\mathbf k_4})
\picube\, \sdelta{\bkone+\bktwo+\bkthree+\bkfour}\,.
\label{zeta4}
\ee
We need twelve real numbers to specify a set of four three-dimensional momenta. However, momentum conservation, $\bkone+\bktwo+\bkthree+\bkfour=0$, eliminates three of them and invariance under rotational symmetry eliminate three others. We are thus left with six independent parameters that specify inequivalent configurations of the tetrahedron formed by the four $\bk$ vectors: we will use the set $\lbrace k_1,k_2,k_3,k_4,k_{12},k_{23} \rbrace$ where $k_{12}=   | \bk_{1}+\bk_{2}  |$ and  $k_{23}=   | \bk_{2}+\bk_{3}  |$. The others $k_{ij}=   | \bk_{i}+\bk_{j}  |$ can then be reexpressed in terms of them as follows: 
\bea
k_{13}&=&k_{24}=\sqrt{k_1^2+k_2^2+k_3^3+k_4^2-k_{12}^2-k_{23}^2}\,, \nn \\
k_{14}&=&k_{23}\,, \qquad k_{34}=k_{12}\,.
\eea
Note that there are geometrical limitations on the parameter space, for instance in the form of triangle inequalities (see \cite{Chen:2009bc} for details).

From the general $\delta N$ expansion (\ref{expansion}), the primordial trispectrum can be evaluated and one finds \cite{Byrnes:2006vq}
\bea
T_\zeta(\bkone,\bktwo,\bkthree,\bkfour) &=&N_AN_BN_CN_D T^{ABCD}(\bkone,\bktwo,\bkthree,\bkfour) \nn \\
& +& N_{A B}N_C N_D N_E \left[ C^{A C}(k_1)B^{BDE}(k_{12},k_3,k_4)+ 11\,\,
\rm{perms}\right] \nn
\\ & +& N_{A B}N_{C D}N_E N_F \left[C^{BD}(k_{13})C^{AE}(k_3)C^{CF}(k_4) +11\,\,
\rm{perms}\right] 
 \nn \\ & +&N_{A B C}N_D N_E N_F
\left[C^{A D}(k_2)C^{B E}(k_3)C^{C F}(k_4)+3\,\,\rm{perms}\right]\,
\label{trispectrum-general}
 \eea
(where we have omitted special configurations of the wavevectors, for instance when any  $\bk$ vector or the sum of any two $\bk$ vectors is zero). Here the $T^{ABCD}$ are the connected four-point correlation functions of the field perturbations at the epoch of horizon crossing, defined by
\bea
&& \hspace*{-1.5em} \la Q^A_{{\mathbf k_1}}\,Q^B_{{\mathbf k_2}}\, Q^C_{{\mathbf k_3}}
Q^D_{{\mathbf k_4}} \ra_c \equiv T^{ABCD}(\bkone,\bktwo,\bkthree,\bkfour) \picube \, \sdelta{\sum_i \bk_i}\,. 
\label{deftrispectrum}
 \nn 
 \eea
Introducing 
\be
  \tau_{NL}\equiv\frac{N_{AB}N^{AC}N^BN_C}{(N_DN^D)^3}\, 
   \label{tauNLmultifield}
  \ee
  and
  \be
  g_{NL}\equiv \frac{25}{54}\frac{N_{ABC}N^AN^BN^C}{(N_DN^D)^3}\,
   \label{gNLmultifield}
 \ee
 simplifies the expression of (\ref{trispectrum-general}) to
\bea
T_\zeta (\bkone,\bktwo,\bkthree,\bkfour)&=&
N_AN_BN_CN_D T^{ABCD}(\bkone,\bktwo,\bkthree,\bkfour)  \nn  \\
&+& N_{A B}N_C N_D N_E \left[ C^{A C}(k_1)B^{BDE}(k_{12},k_3,k_4)+ 11\,\,
\rm{perms}\right]  \nn \\
&+& \tau_{NL}\left[P_\zeta(k_{13})P_\zeta(k_3)P_\zeta(k_4) +11\,\,\rm{perms}\right] \nn \\ 
&+&\frac{54}{25}g_{NL}\left[P_\zeta(k_2)P_\zeta(k_3)P_\zeta(k_4)+3\,\,\rm{perms}\right] \,.
\label{tauNLgNLdefn} 
\eea
In order to compare the various momentum dependences entering (\ref{tauNLgNLdefn}), we use the form factor $\CT$defined by \cite{Chen:2009bc}
\be
T_{\z} (\bkone,\bktwo,\bkthree,\bkfour)  = (2\pi)^6\, \Pz^3 \left( \prod_{i=1}^4 \frac{1}{k_i^3} \right) ~ \CT(k_1,k_2,k_3,k_4,k_{12},k_{14})\,,
\ee
whilst the amplitude of the trispectrum signal is quantified by the estimator $t_{NL}$ for each shape component 
\bea
\frac{1}{k^3} \CT(k_1,k_2,k_3,k_4,k_{12},k_{14})_{\rm component} \xrightarrow[\rm limit]{\rm RT} \tNL\,.
\eea
Here the regular tetrahedron (RT) limit means $k_1=k_2=k_3=k_4=k_{12}=k_{14}\equiv k$ \cite{Chen:2009bc,Gao:2009at}.

In the DBI regime of brane inflation, the quantum non-Gaussianities of the fields are large -- as set by their bispectrum $B^{ABC}$ and connected part of their trispectrum $T^{ABCD}$ -- and the first two lines in Eq.~(\ref{tauNLgNLdefn}) a priori gives sizeable contributions to the primordial trispectrum. Moreover, the last two lines in Eq.~(\ref{tauNLgNLdefn}) could be significant as we will see later. Their respective amplitude are determined by the two parameters $\tau_{NL}$ and $g_{NL}$, which are similar to $\Floc$ for the bispectrum in that they describe the nonlinearities generated classically outside the horizon. In that sense we will occasionally refer to them as the local trispectrum parameters. They contribute to the form factor as
\bea
\CT \supset \frac{36}{25} \tau_{NL} T_{loc1} + g_{NL} T_{loc2}\,,
\label{Tlocs}
\eea
where the two local shapes are
\bea
T_{loc1} &=& \frac{9}{50} \left( \frac{k_1^3 k_2^3}{k_{13}^3} +11\,\,\rm{perms} \right) ~,
\label{Tloc1}
\\
T_{loc2} &=&
 \frac{27}{100} \sum_{i=1}^4 k_i^3\,,
\label{Tloc2}
\eea
and the size of the trispectrum for each is
\bea
t_{NL}^{loc1} = 1.18\, \tau_{NL} ~, ~~~
t_{NL}^{loc2} = 1.08 \,g_{NL} ~.
\eea

We study the local and intrinsically quantum contributions to the trispectrum in the next subsection while the discussion of the second term in the right hand side of Eq.~(\ref{tauNLgNLdefn}) is deferred to subsection \ref{new}.

\subsection{Local and intrinsically quantum trispectra in our model}
\label{subsec:loc-quantum}

We begin by considering the standard local parameters $\tau_{NL}$ and $g_{NL}$. From their definitions Eqs.~(\ref{tauNLmultifield}) and (\ref{gNLmultifield}), they are given by
\be
\tau_{NL}=\frac{N_s^2 N_{ss}^2}{\left(N_{\s}^2+N_s^2 \right)^3} \,, \qquad g_{NL}=\frac{25}{54} \frac{N_s^3 N_{sss}}{\left(N_{\s}^2+N_s^2 \right)^3}\,.
\label{tau-g-2field}
\ee
Notice that using Eqs.~(\ref{pe1})-(\ref{pe3}) and (\ref{Ns})-(\ref{Nsss}), the entropic derivatives satisfy
\be
N_s N_{sss}=3 \sin^2(\b) N_{ss}^2 \,.
\ee
We therefore obtain
\be
g_{NL}=\frac{25}{18}\sin^2(\b) \tau_{NL}\,.
\label{g-tau}
\ee
As $\t_{NL}$ and $g_{NL}$ are in principle observationally distinguishable \cite{Okamoto:2002ik}, this gives us the possibility to deduce the angle $\b$ from the observations. Given that $g_{NL} < \frac{25}{18}\, \tau_{NL}$, it should be stressed also that one can not obtain a large $g_{NL}$ without having a large $\tau_{NL}$, contrary to the cases of the curvaton \cite{Sasaki:2006kq}, ekpyrotic models \cite{Lehners:2009ja} or non slow-roll multifield inflation \cite{Byrnes:2009qy}. Specializing to the limit in which the curvature perturbation is mostly of entropic origin, one obtains from Eq.~(\ref{tau-g-2field}) the single-field \textit{entropic} result
\be
\tau_{NL}=\frac{N_{ss}^2}{N_{s}^4}=\left(\frac{6}{5}\Floc \right)^2\,, \qquad \T^2 \gg 1
\label{tau-entropic}
\ee
where $\Floc$ is given in this limit by (\ref{FNL4-iso}) (and the relation (\ref{g-tau}) still holds).

\bigskip

The intrinsic trispectra of the fields also leave their imprint on the primordial trispectrum, through the linear relation between the curvature perturbation and the field perturbations at horizon crossing (see the first term in Eq.~(\ref{tauNLgNLdefn})). The calculation of the field trispectra in multifield models with non-standard kinetic terms, and in multifield DBI inflation in particular, begun only very recently \cite{Gao:2009gd,Mizuno:2009cv,Gao:2009at}. For example, the authors of \cite{Mizuno:2009cv,Gao:2009at} considered the trispectra from the intrinsic fourth-order contact interaction. However, it was pointed out that there are other important contributions coming from interactions at a distance \cite{Seery:2008ax,Gao:2009gd,Chen:2009bc,Arroja:2009pd}. As the full calculation of the field bispectra in multifield DBI inflation has not yet been completed, below we simply comment on the partial results coming from the contact interaction. The important result is that the degeneracy between single and multiple-field models is broken at this level, {\it i.e.} with our notations one obtains
\bea
T_\zeta(\bkone,\bktwo,\bkthree,\bkfour) &\supset & N_{\s}^4 \bigg(T^{\s \s \s \s}(\bkone,\bktwo,\bkthree,\bkfour)+\T^2 \Big( T^{\s \s s s}(\bkone,\bktwo,\bkthree,\bkfour)+5\,\,\rm{perms}\Big)
\cr
&+& \T^4 \, T^{s s s s}(\bkone,\bktwo,\bkthree,\bkfour)  \bigg)
\cr
&=& N_{\s}^4 (1+\T^2) \bigg(T^{\s \s \s \s}(\bkone,\bktwo,\bkthree,\bkfour) + \T^2 T^{s s s s}(\bkone,\bktwo,\bkthree,\bkfour)  \bigg)
\label{tri}
\eea
where the last line follows from a particular relation between the mixed term $T^{\s \s s s}$ and the adiabatic and entropic trispectra, which have the same order of magnitude $\sim N_{\s}^2 H_*^6/\css^4$ but different momentum dependence. Therefore the trispectrum can a priori discriminate between single- and multifield DBI inflation. In particular, one can hope that looking at the trispectrum signal in different limits of the tetrahedron's parameter space might enable a measurement of $\T$. As for the overall amplitude of these terms though, with $\Pz=\left( \frac{H_*}{2 \pi}\right)^2N_{\s}^2(1+\T^2)$ (\ref{curvature-general}), this schematically gives a contribution to $\tNL$ of the form
\be
\tNL \sim \frac{1}{\Pz^3} \frac{N_{\s}^2 H_*^6}{\css^4}N_{\s}^4(1+\T^2)(1+a\, \T^2) \sim \frac{1}{\css^4}(1+\T^2)\frac{1+a\, \T^2}{(1+\T^2)^3}
\ee
where $a$ is a $O(1)$ numerical coefficient. Similarly to the case of the bispectrum, notice that the field trispectra having the same amplitude implies that the transfer function $\T$ solely determines the multiple-field modification to the single-field result $\tNL \sim \frac{1}{\css^4}$ \cite{Huang:2006eh}. Therefore, even if one does not succeed extracting $\T^2$ from measurements of the trispectrum alone -- for example because $\T^2 \gg 1$ and the entropic signal (the last term in (\ref{tri})) dominates over the adiabatic one -- one can in principle combine measurements of the trispectrum and of the bispectrum (\ref{f_NL3}) $\Feq \approx - 0.3 \frac{1}{\css^2(1+\T^2)}$ to determine both $\css$ and $\T$. This illustrates how the study of  higher-order correlation functions can be used to extract information one can not obtain with the power spectrum alone.

\subsection{An observational signature of multifield DBI inflation}
\label{new}

Going back to the general expression Eq.~(\ref{tauNLgNLdefn}), one sees that the intrinsic bispectrum of the fields also contribute to the primordial trispectrum of the curvature perturbation, {\it i.e.} 
 \bea
 T_\zeta(\bkone,\bktwo,\bkthree,\bkfour) &\supset & N_{A B}N_C N_D N_E \left[ C^{A C}(k_1)B^{BDE}(k_{12},k_3,k_4)+ 11\,\,
\rm{perms}\right]\,.
 \eea
Remarkably, this term is solely determined by quantities that already appeared in lower-order correlation functions. In particular, while $N_{AB}$ determine the amplitude of $\Floc$ (\ref{fNLloc}), the intrinsic bispectra of the fields $B^{ABC}$ set the magnitude of $\Fthree$ (\ref{fNLeq}). Therefore, this contribution can be non negligible only if significant classical nonlinearities and quantum non-Gaussianities are present in the same model, which explains why it has been neglected before and why we refer to it as $\Tz$ in the following. Note that since the shape of the field three-point functions does not take a universal form, one cannot extract a number that characterizes the amplitude of $\Tz$ independently of a model, contrary to $\tau_{NL}$ (\ref{tauNLmultifield}), which also depends only on lower-order terms.

In our scenario, $\Tz$ reads
\bea
\Tz&=&  \left[ N_{\s \s} N_{\s}^3 C^{\s \s}(k_1) B^{\s \s \s}(k_{12},k_3,k_4) + N_{\s \s} N_{\s} N_{s}^2 C^{\s \s}(k_1) B^{\s s s}(k_{12},k_3,k_4) 
\right.
\nn
\\
&&\left. 
+ N_{s s} N_{\s} N_{s}^2 C^{s s}(k_1) \left( B^{ s s \s}(k_{12},k_3,k_4)+B^{ s \s s}(k_{12},k_3,k_4)  \right)+ 11\,\,
\rm{perms}\right] \,.
\label{trispectrum-our-case}
 \eea
Notice that the first two terms, being proportional to $N_{\s \s}$, are negligible to leading order in the slow-varying parameters and we are left with the second line alone at this order. We have included the first line only to stress that the trispectrum breaks the degeneracy between the two shapes $b^{\s \s \s}$ and $b^{\s s s}$ entering the bispectrum, defined respectively in Eqs.~(\ref{b-adiabatic}) and (\ref{b-entropic}). Recall indeed that we used the symmetry property
\be
 b^{\s s s}( k_1,k_2,k_3)+ b^{s \s s}( k_1,k_2,k_3)+ b^{s s \s}( k_1,k_2,k_3)= b^{\s \s \s}( k_1,k_2,k_3)
 \ee
 to show that the multifield effects do not modify the shape of $\Fthree$ compared to the single-field case. Here, one sees that because $N_{\s \s} \neq N_{ss}$ in general, one can not use this identity and the primordial trispectrum truly depends on the shape $b^{\s ss}$, not on its symmetrized version $b^{\s \s \s}$.

Let us now concentrate on the leading-order term, namely the second line in Eq.~(\ref{trispectrum-our-case}), and write its contribution to the form factor as
\bea
\CT \supset \sNL \, \Tn \,,
\eea
where
\bea
\Tn& \equiv& \frac{81}{175}\, (k_2 k_3 k_4)^3 \left[ b^{ s s \s}(k_{12},k_3,k_{4})+b^{s \s s}(k_{12},k_3,k_{4})\right]+ 11\,\, \rm{perms}\,.
\label{shape}
\eea
Here, $\sNL$ is a dimensionless non-linearity parameter that set the amplitude of $\Tn$, similarly to $\tau_{NL}$ and $g_{NL}$ for $T_{loc1}$ and $T_{loc2}$ respectively (see Eq.~(\ref{Tlocs})), and the numerical factor in the definition of $\Tn$ is for convenience only. The overal amplitude of this contribution to the trispectrum is 
\be
t_{NL}^{loc, \, eq} = 0.09 \,\sNL ~.
\ee
By noting that (\ref{three-point}) can be rewritten in the form 
\bea 
B^{A B C}(k_1,k_2,k_3) =- \frac{ N_{\s} H_{*}^4}{4 \css^2}  b^{A B C}(k_1,k_2,k_3)\,,
\eea
from (\ref{trispectrum-our-case}), one formally obtains
\bea
\sNL=-\frac{175}{648}\frac{N_{ss}}{N_{\s}^2}\frac{1}{\css^2}\frac{\T^2}{(1+\T^2)^3}\,,
\eea
which is exactly the product of $\Floc$ (\ref{Floc-general}) and $\Feq$ (\ref{f_NL3})
 \be
\sNL =\Floc \, \Feq\,.
 \label{imprint}
 \ee
It is worth emphasizing that, in order to derive the result (\ref{imprint}), we solely used the $\delta N$ formalism together with the general results for the field bispectra \cite{Langlois:2008qf}. Hence it is clear that it is valid for every DBI model, not restricting to a radial trajectory nor to a specific scenario for the entropy to curvature transfer. In particular, we considered the two-field case for simplicity of presentation but the proof generalizes straightforwardly to a higher number of light fields. The consistency relation (\ref{imprint}) is thus structural to multifield DBI inflation. It represents a distinct observational imprint on the trispectrum of the presence of light scalar fields, other than the inflaton, and with non standard kinetic terms, hence the appearance of both the local and equilateral non-linearity parameters.

\subsection{Shapes of trispectra}

\begin{figure}
  \center
  \includegraphics[width=0.6\textwidth]{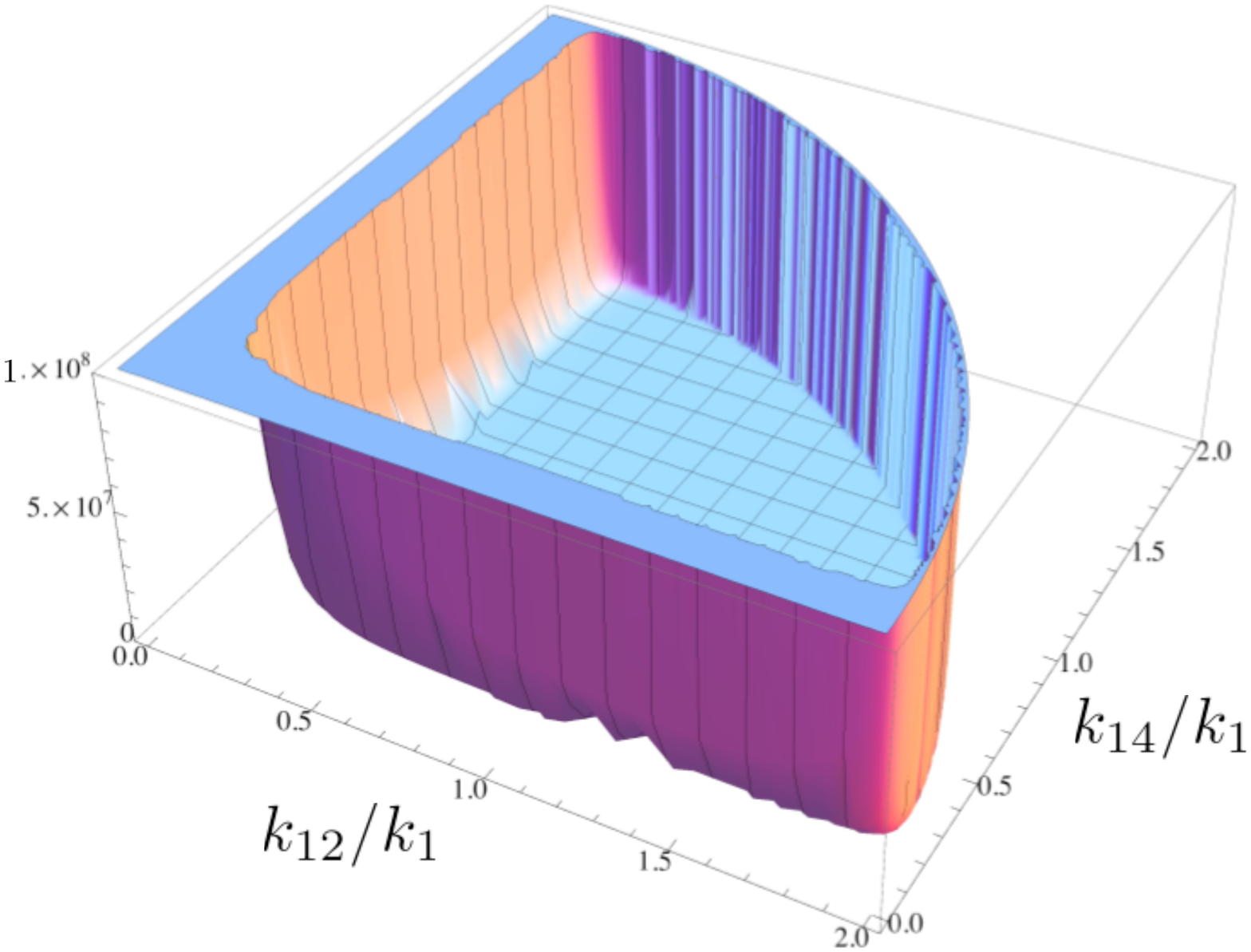}
   \includegraphics[width=0.6\textwidth]{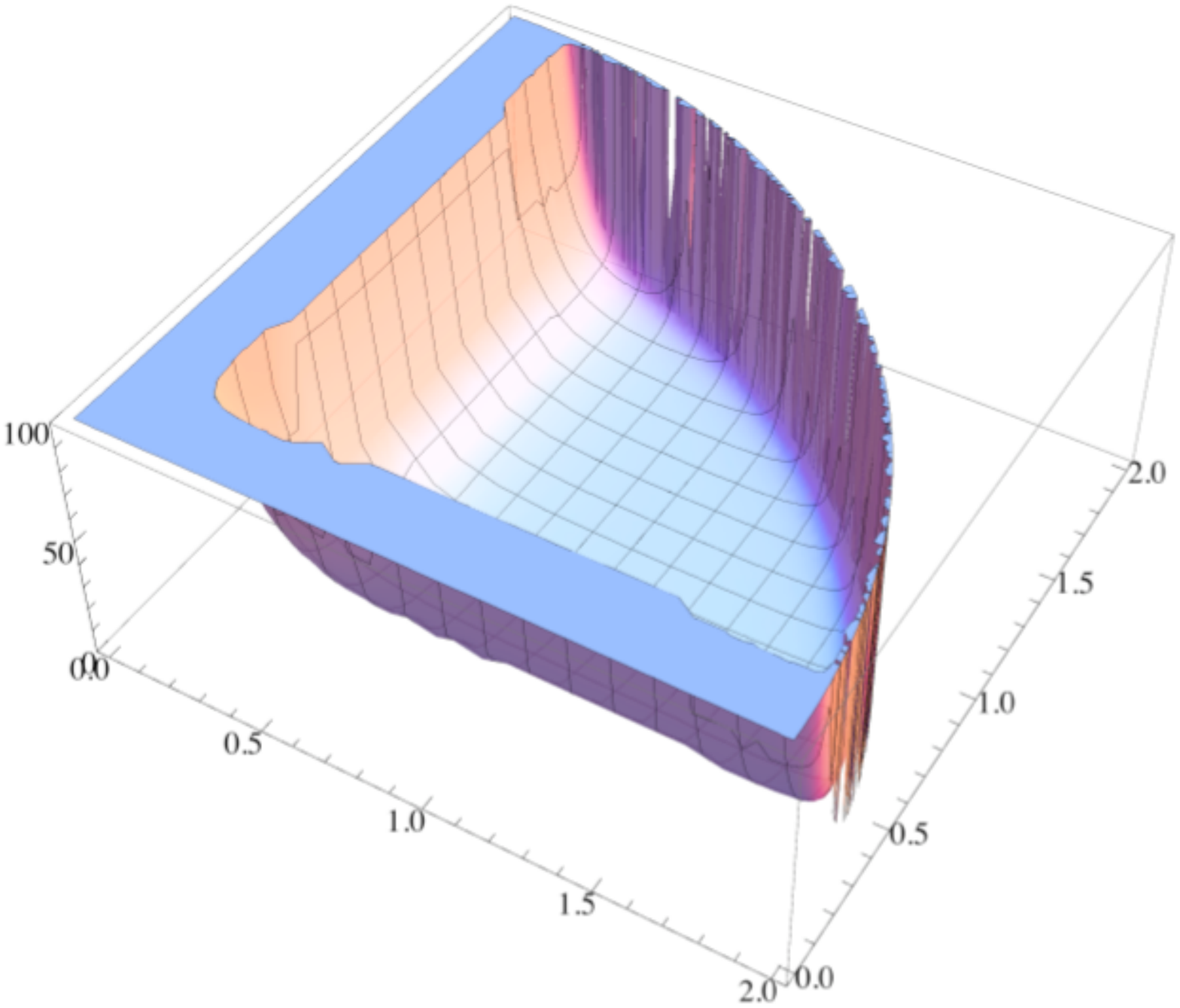}
   \includegraphics[width=0.6\textwidth]{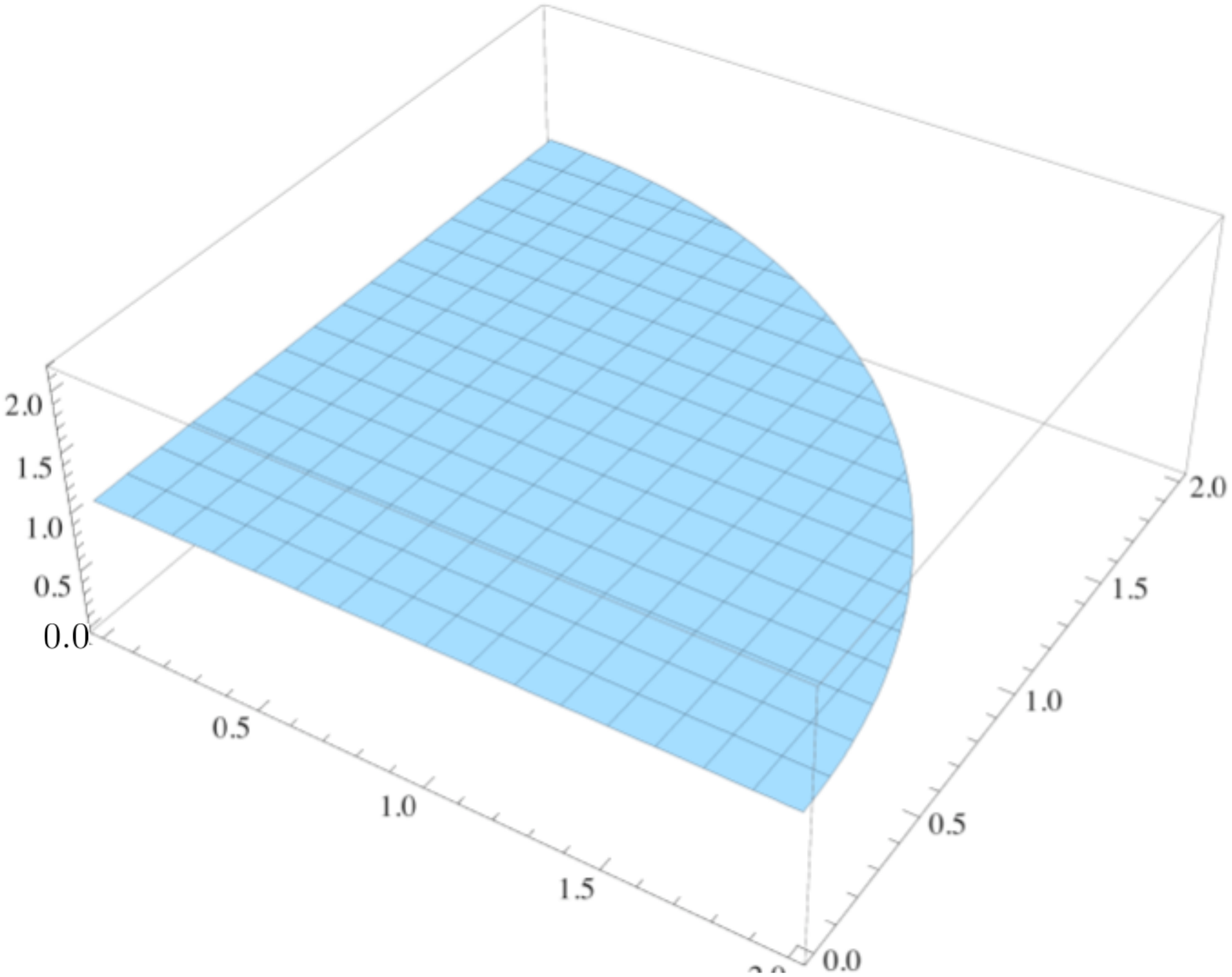}
    \caption{\label{equilateral} In this group of figures, we consider
      the equilateral limit $k_1=k_2=k_3=k_4$, and plot $\Tn$, 
      $T_{loc1}$ and $T_{loc2}$,
      respectively, as
      functions of $k_{12}/k_1$ and
      $k_{14}/k_1$. Note that
    $\Tn$ and $T_{loc1}$ blow up when $k_{12}\ll k_1$ and $k_{14} \ll
    k_1$, as well as in the other boundary, corresponding to $k_{13}\ll k_1$.}
\end{figure}

\begin{figure}
  \begin{center}
   \includegraphics[width=0.6\textwidth]{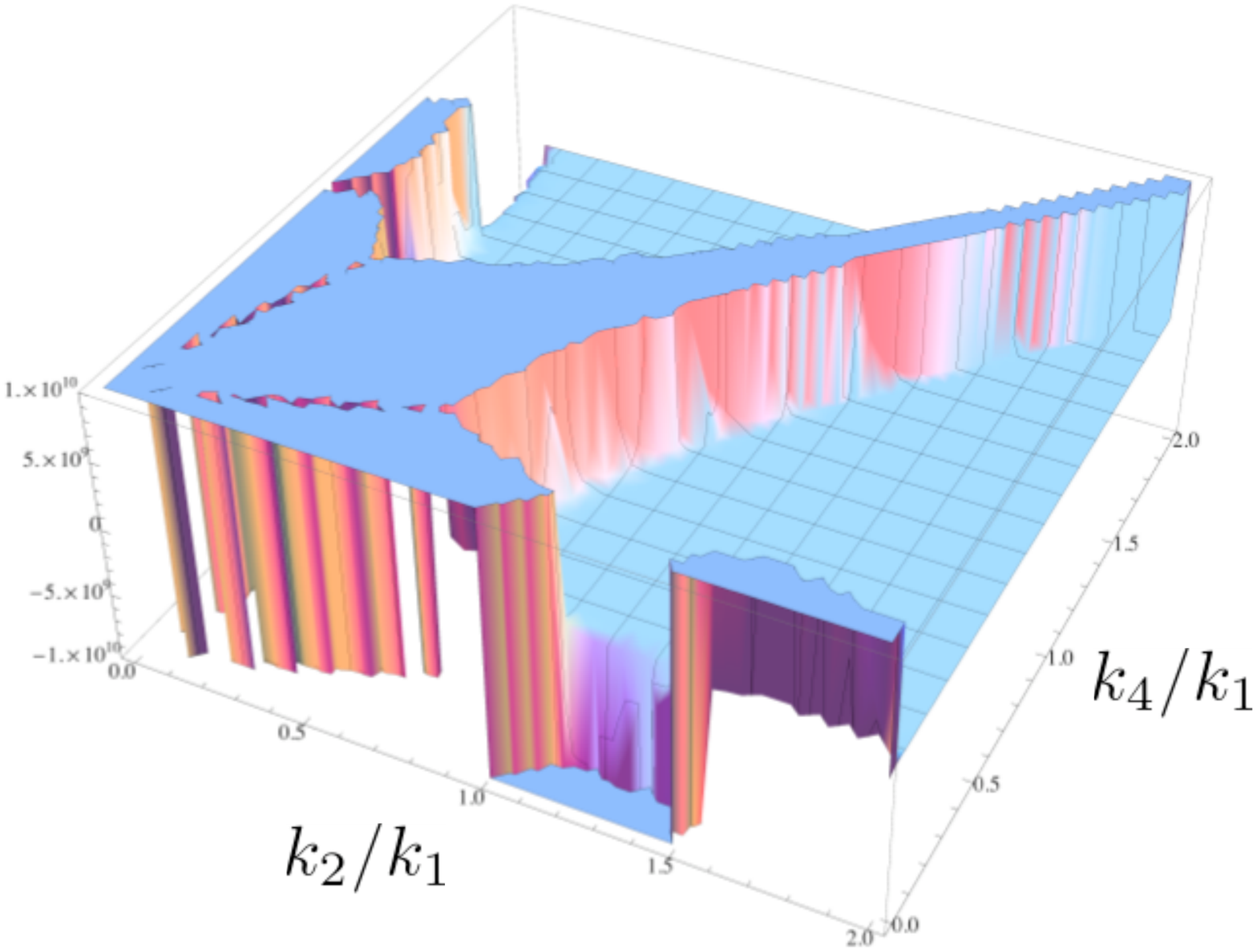}
 \includegraphics[width=0.6\textwidth]{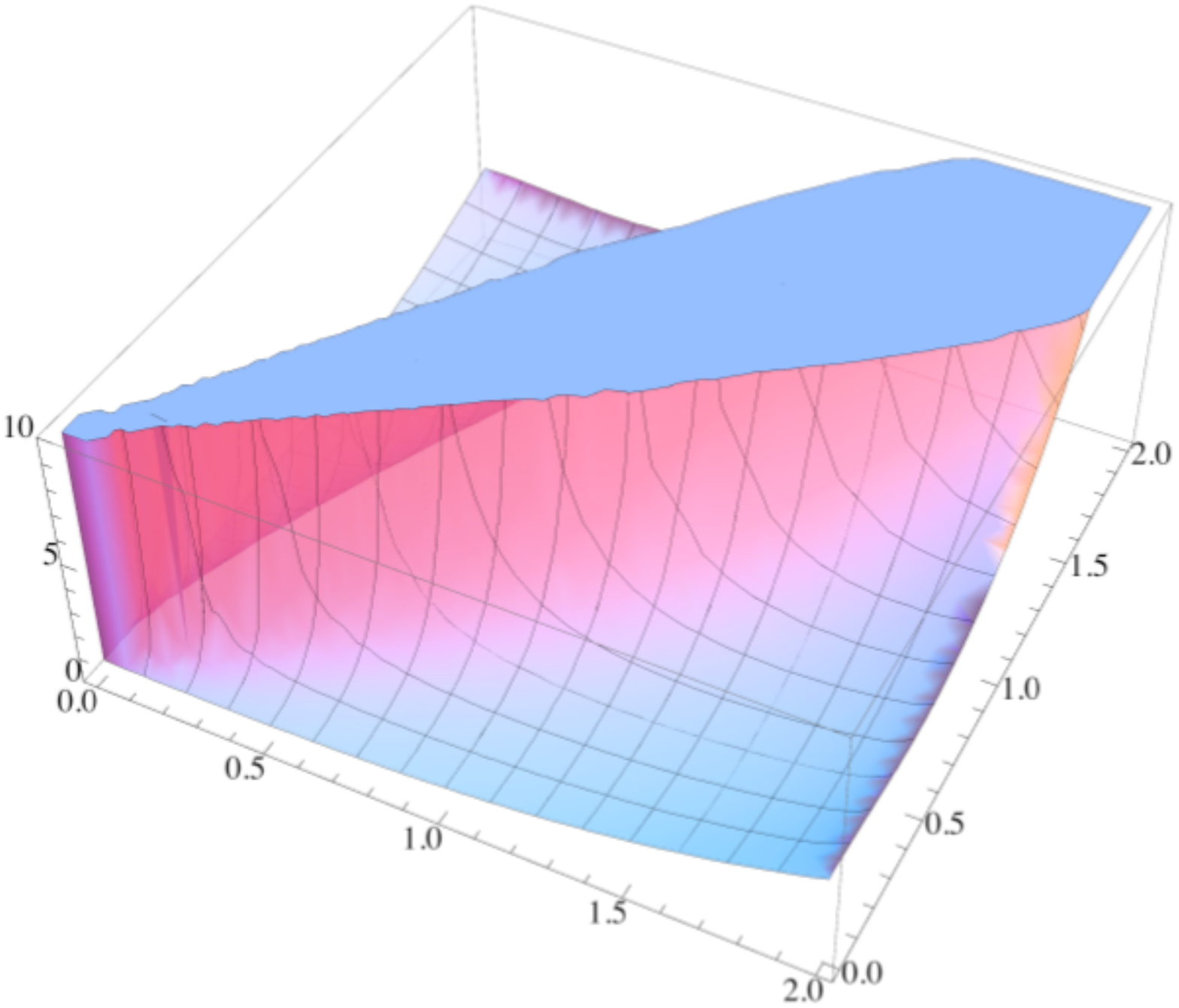}
 \includegraphics[width=0.6\textwidth]{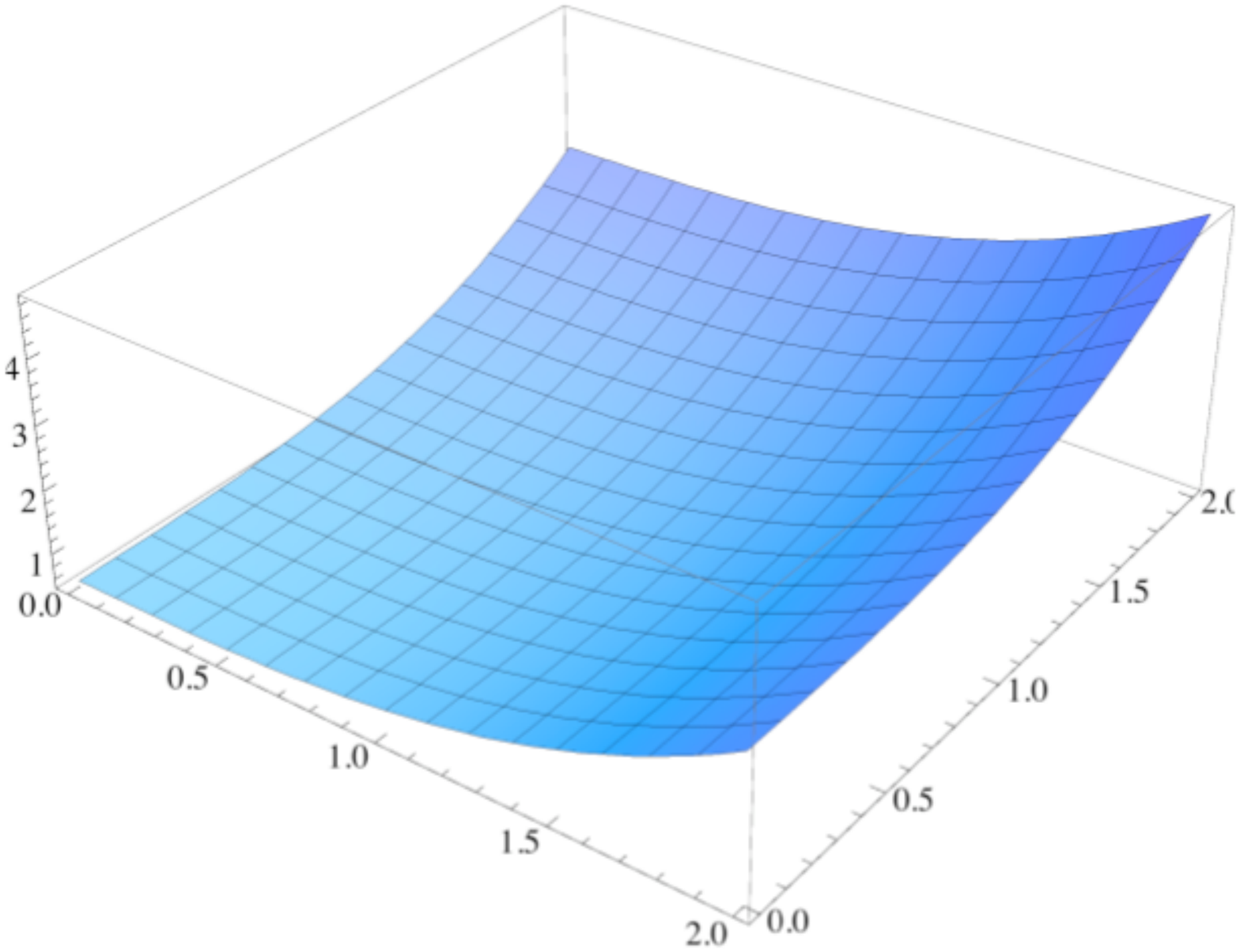}
     \caption{\label{specplanar} In this group of figures, we consider
      the specialized planar limit with $k_1=k_3=k_{14}$, and plot
      $\Tn$, $T_{loc1}$ and
      $T_{loc2}$, respectively, as
      functions of $k_{2}/k_1$ and
    $k_{4}/k_1$. Along the diagonal $k_2 \rightarrow k_4$, $\Tn$ and 
    $T_{loc1}$ blow up because in
this limit, $k_{13}\rightarrow 0$.  At the $k_2\rightarrow 0$ and $k_4\rightarrow
0$ boundaries, $\Tn$ blow up while the local shapes remain finite. Notice that the sign of $\Tn$ varies non trivially over the parameter space.
 }
  \end{center}
\end{figure}

\begin{figure}
  \center
  \includegraphics[width=0.6\textwidth]{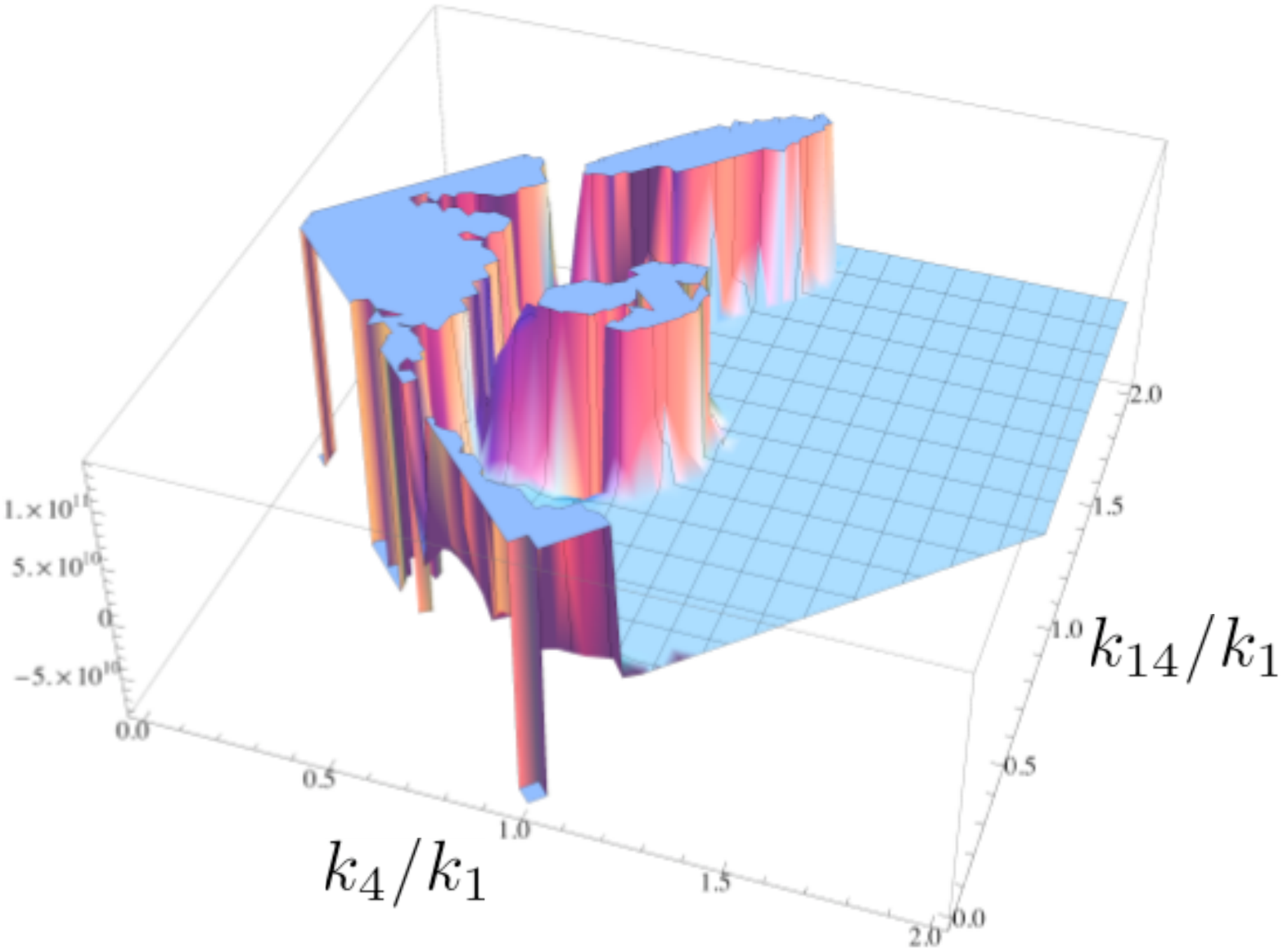}
  \includegraphics[width=0.6\textwidth]{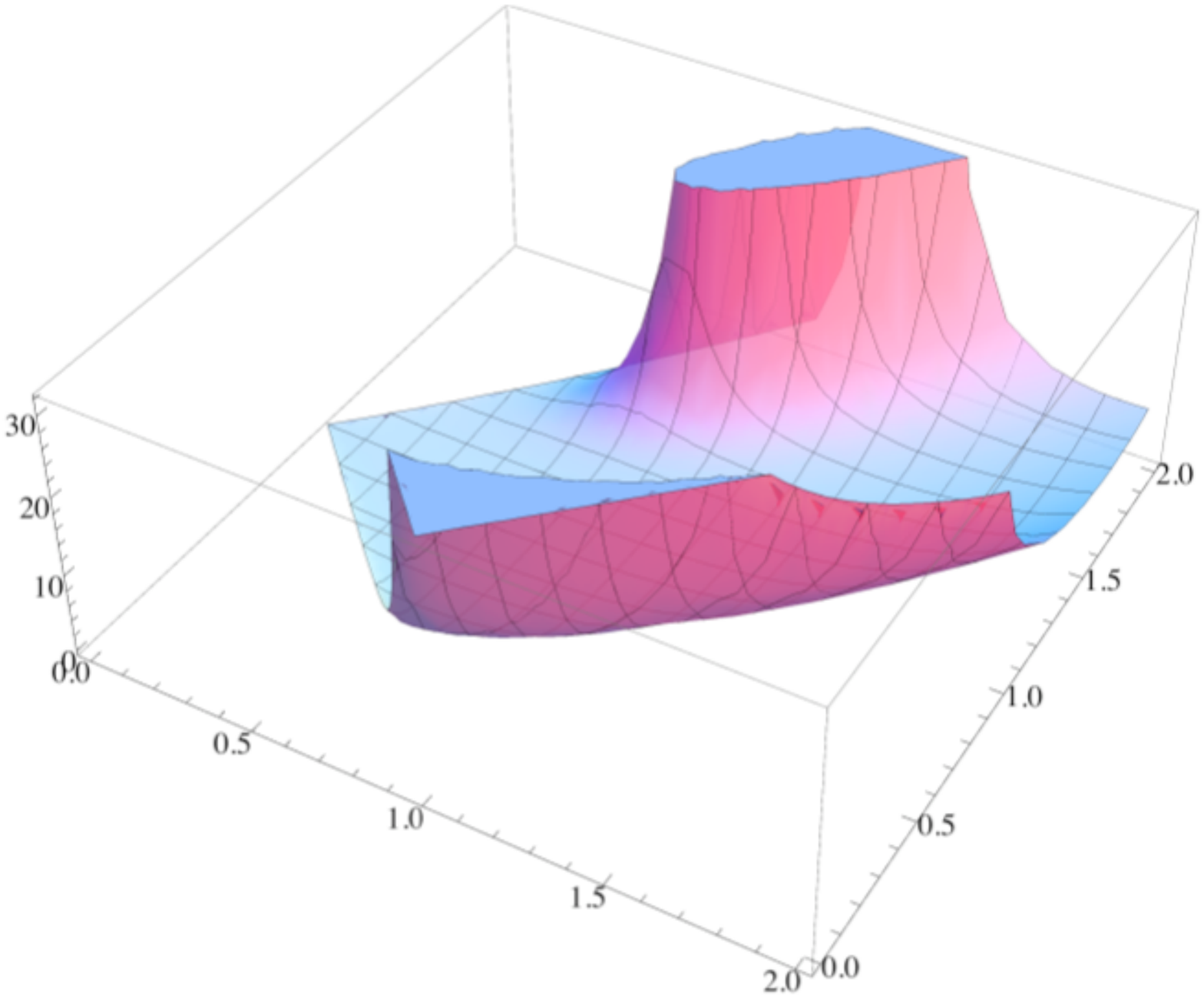}
 \includegraphics[width=0.6\textwidth]{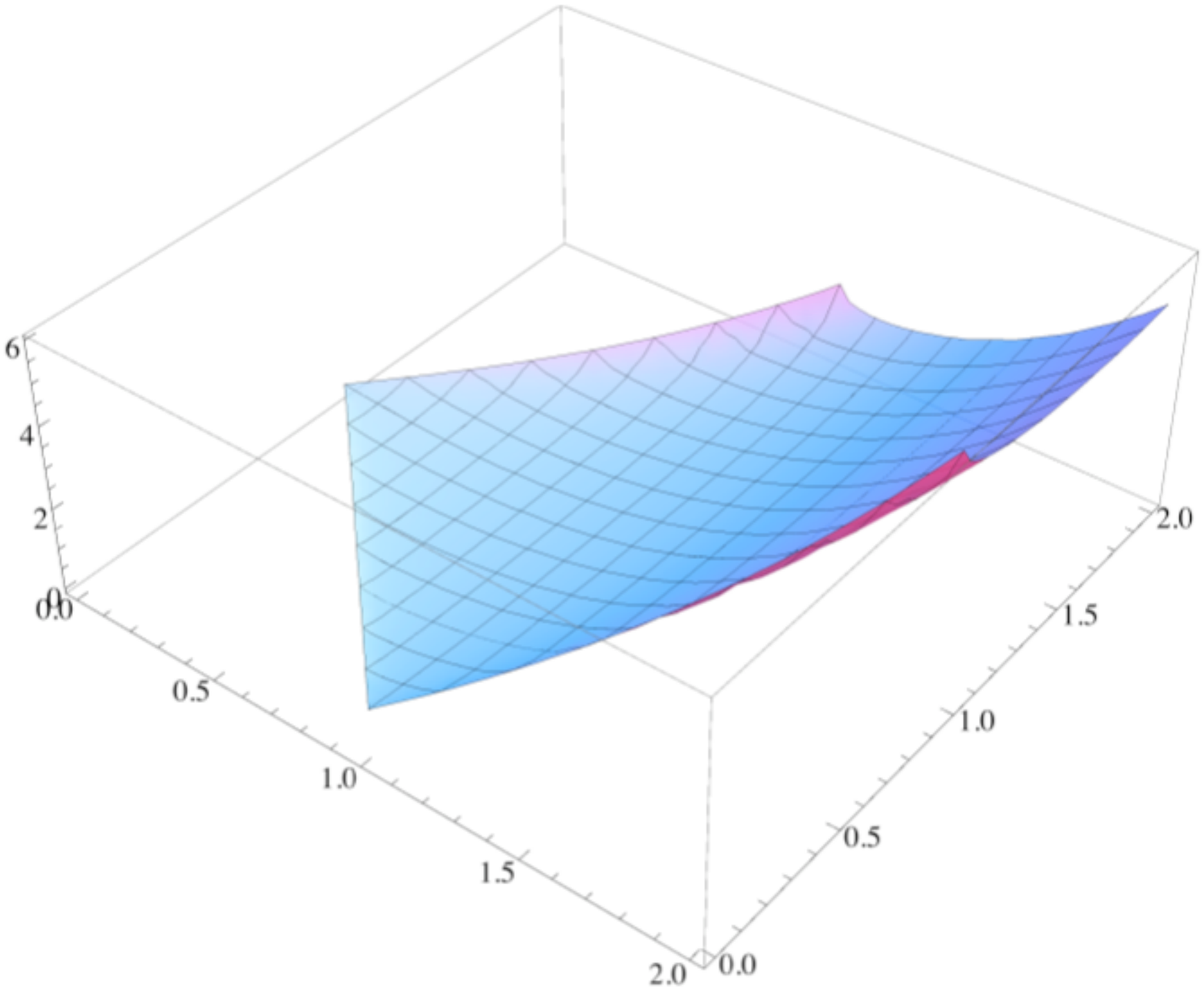}
    \caption{\label{doublesqueeze} In this group of figures, we look
      at the shapes near
      the double squeezed limit: we consider the case where
      ${k}_3={k}_4=k_{12}$ and
      the tetrahedron
  is a planar quadrangle. We plot
      $\Tn$, $T_{loc1}$ and $T_{loc2}$,
      respectively, as
      functions of $k_{4}/k_1$ and $k_{14}/k_1$. Notice that the sign of $\Tn$ varies non trivially over the domain. In the squeezed limit, at
  $(k_4/k_1=1,k_{14}/k_1=1)$ where $k_2\to 0$, and in the double-squeezed
  limit, $k_3=k_4\rightarrow
  0$, $\Tn$ blows up while the local shapes are finite (see the main text for additional comments).      
}
\end{figure}

To test the consistency relation (\ref{imprint}), one must observationally disentangle the contribution proportional to $\sNL$ from other components of the trispectrum. The study of its six-dimensional parameter space is challenging and in the following, we follow the discussion in \cite{Chen:2009bc} and simply plot the form factor $\Tn$ in various limits to reduce the number of variables, and compare it to $T_{loc1}$ and $T_{loc2}$. The shape functions are left white when the momenta do not form a tetrahedron. We consider the following cases (see \cite{Chen:2009bc} for details):

\begin{enumerate}
  \item Equilateral limit: $k_1=k_2=k_3=k_4$. In Fig. \ref{equilateral}, we
  plot $\Tn$, $T_{loc1}$ and $T_{loc2}$ as
  functions of $k_{12}/k_1$ and $k_{14}/k_1$. 
  One observes that $\Tn$ and $T_{loc1}$ blow up at all boundaries, corresponding to  $k_{12}/k_1$, $k_{14}/k_1$ and $k_{13}/k_1$ $\rightarrow 0$. However, the signal in $\Tn$ is much more important, as is clear from the extremely large cutoff for the z-axis in this figure. On the contrary, $T_{loc2}$ is constant in this region of parameter space.

  \item Specialized planar limit: we take $k_1=k_3=k_{14}$, and the
    tetrahedron to be a planar quadrangle with \cite{Chen:2009bc} 
\begin{align}
      k_{12}=\left[
k_1^2+\frac{k_2 k_4}{2 k_1^2}\left( k_2 k_4 +
\sqrt{(4k_1^2-k_2^2)(4k_1^2-k_4^2)} \right) \right]^{1/2}~.
    \end{align}
     We plot the shape functions as functions
of $k_2/k_1$ and $k_4/k_1$ in Fig. \ref{specplanar}. 
At the $k_2 \rightarrow k_4$
limit, we have $k_{13} \to 0$, so that $\Tn$ and $T_{loc1}$ blow up. Again, the signal in $\Tn$ is much more important and we had to use an extremely large cutoff for the z-axis for the sake of presentation. The most distinctive difference comes from the $k_2\rightarrow 0$ and $k_4\rightarrow
0$ boundaries, where $T_{loc1}$ and $T_{loc2}$ take finite (non-zero) values whereas $\Tn$ blows up. Moreover, its sign alternates non-trivially over the parameter space. Interestingly, this feature was not present in any other shape studied in \cite{Chen:2009bc}.

\item Near the double-squeezed limit: we consider the case where
  ${k}_3={k}_4=k_{12}$ and the tetrahedron
  is a planar quadrangle with \cite{Chen:2009bc} 
\begin{align}\label{planark2}
   k_2= \frac{\sqrt{k_1^2 \left(-k_{12}^2+k_3^2+k_4^2\right)- k_{s1}^2 k_{s2}^2+k_{12}^2 k_{14}^2+k_{12}^2 k_4^2+k_{14}^2
   k_4^2-k_{14}^2 k_3^2-k_4^4+k_3^2 k_4^2}}{\sqrt{2} k_4}~,
  \end{align}
where $k_{s1}$ and $k_{s2}$ are defined as
\begin{align}
&  k_{s1}^2\equiv 2\sqrt{(k_1 k_4+{\bf k}_1 \cdot {\bf k}_4)(k_1
k_4-{\bf k}_1 \cdot {\bf
  k}_4)}~,\nonumber\\ &
k_{s2}^2\equiv 2\sqrt{(k_3 k_4+{\bf k}_3 \cdot {\bf k}_4)(k_3
k_4-{\bf k}_3 \cdot {\bf
  k}_4)}~.
\end{align}
  We plot $\Tn$, $T_{loc1}$
  and $T_{loc2}$ as
  functions of $k_{4}/k_1$ and $k_{14}/k_1$ in
  Fig. \ref{doublesqueeze}.  Similarly to the specialized planar limit, notice that the sign of $\Tn$ varies non trivially over the domain. Several differences between the shapes are visible. 1) In the squeezed limit, at
  $(k_4/k_1=1,k_{14}/k_1=1)$ where $k_2\to 0$, and in the double-squeezed
  limit, $k_3=k_4\rightarrow
  0$, $\Tn$ blows up while the local shapes are finite. 2) In the folded limit $(k_4/k_1=1,k_{14}/k_1=0)$, both $\Tn$ and $T_{loc1}$ blow up and $T_{loc2}$ takes a finite value. 3) In the other folded limit, $(k_4/k_1=1,k_{14}/k_1=2)$, all shapes remain finite. Close to it, note that the bump in $T_{loc1}$ actually remains finite while $\Tn$ truly blows up.
       \end{enumerate}

To summarize, although we did not make an exhaustive study of the shape factor $\Tn$, we have seen that it presents characteristic features. In particular, in the last two envisaged situations where the tetrahedron reduces to a planar quadrangle, its sign non-trivially varies over the parameter space, in contrast with all shapes studied in \cite{Chen:2009bc} for instance. This is all the more interesting as this planar limit corresponds to the situation effectively probed by small (angular) scale CMB experiments. This gives hope to be able to test the consistency relation (\ref{imprint}).

\section{Conclusions}
\label{sec:discussion}

In this paper, we have analyzed a scenario to convert entropic into curvature perturbations in the context of multifield brane inflation, and multifield DBI inflation in particular. Considering a standard single-field driven inflationary phase ending by tachyonic instability, we have shown how the perturbations of light fields, parametrizing the angular position of the brane in the throat, can generate curvature perturbation by spatially modulating the duration of inflation. The entropic transfer generated by this mechanism is more efficient in the DBI than in the slow-roll regime, being enhanced by the inverse of the sound speed at horizon crossing and at the end of inflation.

We have investigated the non-Gaussianities created in such models. Single-field DBI inflation is known to produce a substantial amount of non-Gaussianities of equilateral type and previous works on multifield DBI inflation showed that the entropic perturbations reduce the amplitude $\Feq$ of these non-Gaussianities. However, as soon as light scalar fields other than the inflaton are present, non-Gaussianities of another shape, namely local non-Gaussianities, can be present. In our model, they arise because of the nonlinear relation between the curvature perturbation and the angular ones. This local contribution $\Floc$ to the non-linearity parameter $\fNL$ can be large and even saturate the observational bounds for some specific scenarios. 

We have also calculated $\tau_{NL}$ and $g_{NL}$, which are trispectrum parameters similar to $\Floc$ for the bispectrum, in that they describe the nonlinearities generated classically outside the horizon. We have shown that $g_{NL}$ is related to $\tau_{NL}$ by a simple parameter describing the angular position of the mobile brane, implying $g_{NL} < \frac{25}{18}\,\tau_{NL}\,$. When the curvature perturbation is mostly of entropic origin, $\tau_{NL}=\left(\frac{6}{5}\Floc \right)^2$, easily reaching the expected Planck sensitivity $  | \tau_{NL} |  \sim 560 $ \cite{Kogo:2006kh} as soon as $\Floc \gtrsim 25$. The trispectrum also contains a part that is directly related to the trispectra of the field perturbations. One can in principle extract the entropic transfer function from its complicated momentum dependence, as well as determine the sound speed when the observable modes cross the horizon by combining it with measurements of $\Feq$.

Finally, due to the presence of light scalar fields, other than the inflaton, and with non standard kinetic terms, the primordial trispectrum acquires a particular momentum dependent component whose amplitude (\ref{imprint}) is given by the product $\Floc\, \Feq$. Moreover, this consistency relation is valid in every multifield DBI model, whatever the brane's trajectory or the mechanism to convert entropic perturbations into the curvature perturbation. It thus represents an interesting observational signature of multifield DBI inflation. We have represented the corresponding form factor in different limits and it turned out to display important differences with the shapes associated to $\tau_{NL}$ and $g_{NL}$, such as its sign varying over the tetrahedron's parameter space. Overall, should future experiments come to detect both local and equilateral non-Gaussianities in the primordial bispectrum, the trispectrum may well help to to confirm or exclude this kind of scenarios.

\emph{Note added}: On the day this work was submitted, the paper \cite{Mizuno:2009mv} appeared in the arXiv, which completes the calculation of the trispectrum coming from the four-point functions of the field perturbations in multifield DBI inflation.

\vspace{1cm}

\noindent {\it Acknowledgements} 

We would like to thank Eiichiro Komatsu, David Langlois, Liam McAllister, Fransesco Nitti, Sarah Shandera and Daniele Steer for valuable discussions related to the topic of this paper, and particularly David Langlois and Daniele Steer for their careful reading of the manuscript.  

\vspace{1cm}

\section*{Appendix 1 - $\delta N$ formulae beyond leading order}
\label{slow-varying}

Using the expressions of $\zeta_*$ (\ref{zeta*}) and $\zeta_e$ (\ref{zetae}) together with the definitions of $\eta$ (\ref{eta}) and $s$ (\ref{s}), one obtains the coefficients of the $\delta N$ expansion without restricting to leading order in these slowly-varying parameters and their time derivatives:

\bea
&N_{\s}&=- \left.  \frac{1}{\sqrt{2 \e c_{s}}}\frac{1}{\MP} \right|_{*}  \\
&N_s&=- \left. \frac{\p_e' }{\sqrt{2 \e c_{s}}}\frac{1}{\MP}  \right|_{e}   
\eea
\bea
&N_{\s \s}&= \left. \frac{\eta+s}{4 \e c_s}\frac{1}{\MP^2}   \right|_{*}  \label{Nsisi}  \\
&N_{ss}&=-\left. \left( \frac{\p_e^{(2)}}{\sqrt{2 \e c_s}\MP}+\frac{\eta+s}{4 \e c_s}\frac{\p_e'^2}{\MP^2}  \right)    \right|_{e} 
\eea
\bea
&N_{\s \s \s}&= \left. \frac{1}{2^{5/2} (\e c_s)^{3/2}\MP^3}\left( \frac{\dot \eta}{H} +\frac{\dot s}{H}-(\eta+s)^2\right)   \right|_{*}  \\
&N_{sss}&= - \left( \frac{\p_e^{(3)}}{\sqrt{2 \e c_s}\MP}+3\frac{(\eta+s)}{4 \e c_s}\frac{\p_e' \p_e^{(2)}}{\MP^2} 
\right.
\nn
\\
&& \left.
+ \frac{\p_e'^3}{2^{5/2} (\e c_s)^{3/2}\MP^3}  \left. \left( \frac{\dot \eta}{H} +\frac{\dot s}{H}-(\eta+s)^2\right)  \right) \right|_{e} 
\eea

The derivatives of $\p_e(\tht_*)$ (\ref{pe}) are evaluated at the background value $\bar{\tht} =b_* c_{s*} \bar \th$ and their explicit expressions read
\bea
&\p_e'&=-\tan(\b)\frac{\rb}{c_{s*}} \,, \label{pe1} \\
&\p_e^{(2)}&=-\frac{1}{\cos^3(\b)} \left(\frac{\rb}{c_{s*}}\right)^2 \frac{1}{\pc} \,,  \label{pe2} \\
&\p_e^{(3)}&=-\frac{3 \tan(\b)}{\cos^4(\b)} \left(\frac{\rb}{c_{s*}}\right)^3 \frac{1}{\pc^2}  \label{pe3} \,.
\eea

\section*{Appendix 2 - Higher order loop contributions}
\label{loop}

Given we encountered large entropic derivatives in the $\delta N$ formalism, one must ensure that we are in a simple perturbative regime in which one can safely neglect higher order loop corrections (see \cite{Cheung:2007st,Leblond:2008gg} for discussions on the viability of a perturbative expansion and \cite{Byrnes:2007tm} for a diagrammatic approach to loop effects). The integral over the
loop momenta give rise to logarithmic infrared divergences $\ln(kL)$ where $L$ is the large
scale cut off. In the following we assume that one can take $\ln(kL)\sim1$ as done in \cite{Boubekeur:2005fj,Cogollo:2008bi,Rodriguez:2008hy}. Concentrating on the $\delta N$ expansion (\ref{expansion}) to second order for simplicity, and neglecting for the moment the non-Gaussianities of the fields at horizon crossing, the dominant loop corrections to the power spectrum,
bispectrum and trispectrum are \cite{Cogollo:2008bi,Rodriguez:2008hy}
\bea
\Prat&=&\frac{N_{AB}N_{AB}}{(N_C N_C)^2}\Pz, \label{loop-P}  \\
\fNLl&=&\frac{5}{6}\frac{N_{AB}N_{BC}N_{AC}}{(N_D N_D)^3}\Pz, \label{loop-3} \\  
\tau_{NL}^{1\,\mathrm{loop}}&=& \frac{N_{AB}N_{BC}N_{CD}N_{AD}}{\left(N_E N_E\right)^4}\Pz. 
\label{tauNL}
\eea

Below we discuss the two limiting cases of a large and small entropic transfer.
  
\begin{itemize}
  \item \textit{Large entropic transfer.} 
  
For a large entropic transfer, one obtains
\bea
\Prat&=& \left( \frac{N_{ss}}{N_s^2} \right)^2\Pz= \left(\frac{6}{5}\Floc\right)^2 \Pz,  \\
\fNLl&=&\frac{5}{6}  \left( \frac{N_{ss}}{N_s^2} \right)^3 \Pz=\frac{36}{25}  \left(\Floc\right)^3 \Pz, \\  
\tau_{NL}^{1\,\mathrm{loop}}&=&  \left( \frac{N_{ss}}{N_s^2} \right)^4  \Pz= \left(\frac{6}{5}\Floc\right)^4 \Pz.
\eea
Therefore, given that $\Pz\sim10^{-10}$ and that observations already show that the level of non-Gaussianities is relatively small, $\Floc = O(100)$, this demonstrates that loop corrections can be neglected, and clearly the argument is valid for every scenario in which one Gaussian field only is responsible for the curvature perturbation \cite{Lyth:2007jh}.

In general, the field non-Gaussianities lead to additional loop contributions, for instance to $\fNL$. This was shown to be completely negligible for slow-roll models \cite{Zaballa:2006pv} but in DBI inflation where intrinsic non-Gaussianities are large, one has to determine if this remains true. In particular, there is a contribution to the primordial trispectrum $B_\zeta$ (\ref{3-point}) from terms of the form
\be
N_A N_{BC} N_{DE} \la  Q^A(\bkone)  (Q^B \star Q^C)(\bktwo)(Q^D \star Q^E)(\bkthree) \ra \,,
\label{new-loops}
\ee
where the symbol $\star$ denotes a convolution product. The most dangerous terms in (\ref{new-loops}) involve the entropic derivatives and, as to leading order there is no purely entropic three point function, one is led to consider
\be
N_{\s} N_{ss}^2  \la  Q_{\s}(\bkone)  (Q_s \star Q_s)(\bktwo)(Q_s \star Q_s)(\bkthree) \ra \,.
\label{new-loops-important}
\ee
From the definition of $\fNL$ (\ref{fNL-def})  as well as the result (\ref{three-point}) for the three point functions of the fields, its contribution is of order
\be
\fNLl \supset \left( N_{\s} N_{ss}^2 H_*^2   \frac{H_*^4}{ \sqrt{\es \css }c_{s*}^2 \MP}  \right)  \big/ \bigg( H_*^4 N_{\s}^4 \left(1+\T^2\right)^2 \bigg)\,.
\ee
For a large entropic transfer, one therefore obtains
\be
\fNLl \supset f_{NL}^{eq} (f_{NL}^{loc})^2 \Pz \,,
\ee
where we have used (\ref{f_NL3}) and (\ref{Floc-general}). Again, the observational bounds $ f_{NL}^{eq},f_{NL}^{loc} =O(100)$ as well as the normalization $\Pz\sim10^{-10}$ show that such a loop-correction is negligible and one can verify that the same conclusion applies to other higher order corrections.
 \item \textit{Small entropic transfer}

When the entropic transfer is inefficient -- $\T^2 \ll 1$ -- the power spectrum (\ref{curvature}) as well as the equilateral non-Gaussianities (\ref{f_NL3}) remain the same as in the single-field case, so that the multifield aspects we are taking into account can not cure the model if it is under pressure at these levels. As for the local non-Gaussianity parameter, from (\ref{Floc-general}) one obtains
\be
\Floc=\frac{5}{6} \T^2 \frac{N_{s s}}{N_{\s}^2}\,, \qquad \T^2 \ll 1\,.
\ee
Its suppression by the small entropic transfer gives few hope to have a significant $\Floc$, rendering the situation effectively indistinguishable from the single-field case for these observables, unless we have an extremely large ratio $N_{ss}/N_{\s}^2$. For instance, $\T^2 \sim 10^{-2}$ and $N_{ss}/N_{\s}^2 \sim 10^3$ give $\Floc \sim 10$. However, if $N_{ss}/N_{\s}^2$ is made this large, Eqs.~(\ref{loop-3}) and (\ref{tauNL}) give
\bea
\fNLl&=&\frac{5}{6}  \left( \frac{N_{ss}}{N_{\s}^2} \right)^3 \Pz \sim 1\,, \\
\tau_{NL}^{1\,\mathrm{loop}}&=&  \left( \frac{N_{ss}}{N_{\s}^2} \right)^4  \Pz \sim 10^3\,,
\eea
and loop effects become important, which requires a more refined study.

\end{itemize}

\bibliography{biblio}

\end{document}